\documentclass[a4paper,11pt]{article}
\pdfoutput=1 

\usepackage{jcappub} 

\usepackage[T1]{fontenc} 
\usepackage{booktabs}
\usepackage{subcaption}
\usepackage{enumitem} 
\usepackage{multirow}   
\usepackage{makecell}   

\graphicspath{{./plots/}}

\newcommand{\revise}{}

\title{Forecasts of CMB $E$-mode anomalies for AliCPT-1}
\author[a,b,c]{Jiazheng Dou,}
\author[b,c]{Wen Zhao}
\affiliation[a]{Key Laboratory of Particle Astrophysics, Institute of High Energy Physics, Chinese Academy of Sciences, Beijing 100049, People’s Republic of China}
\affiliation[b]{Department of Astronomy, University of Science and Technology of China, Chinese Academy of Sciences, Hefei, Anhui 230026, People's Republic of China}
\affiliation[c]{School of Astronomy and Space Sciences, University of Science and Technology of China, Hefei 230026, People's Republic of China}

\emailAdd{doujzh@mail.ustc.edu.cn}
\emailAdd{wzhao7@ustc.edu.cn}

\abstract{The standard $\Lambda$CDM model has been highly successful in describing cosmic microwave background (CMB) observations. Nevertheless, a set of large-scale statistical anomalies persists in temperature anisotropies across \textit{WMAP} and \textit{Planck}. CMB $E$-mode polarization offers an independent probe of these anomalies, circumventing the look-elsewhere effect inherent in temperature-only analyses. In this paper, we forecast the capability of the Ali CMB Polarization Telescope (AliCPT), a ground-based CMB experiment in the Northern Hemisphere, to detect such anomalies in large-scale $E$-mode polarization. Using 1000 unconstrained simulations processed with the NILC component separation method, we evaluate four anomaly estimators: dipole modulation, lack of large-angle correlations, quadrupole-octopole alignment, and point-parity asymmetry. Our analysis considers two noise levels for AliCPT, \revise{the goal configuration of the Simons Observatory (SO) Large Aperture Telescope (LAT) alone, and a joint AliCPT+SO configuration}. For dipole modulation, we validate the local variance estimator on modulated simulations with an input amplitude $A_d = 0.07$, and find that the combined AliCPT+SO dataset is likely to detect the injected $E$-mode modulation at a 99\% confidence level. Tests of the full suite of anomaly statistics on unconstrained isotropic simulations indicate that AliCPT alone, owing to its limited sky coverage, might introduce systematic biases or enlarged uncertainties, especially for quadrupole-octopole alignment and point-parity asymmetry. The combination with SO largely restores the statistical distributions to those expected in an ideal full-sky scenario, thereby establishing a near-cosmic-variance benchmark for upcoming anomaly investigations.
}


\begin{document}
\maketitle
\flushbottom

\section{Introduction}           
\label{sec:intro}

The standard $\Lambda$CDM paradigm has demonstrated remarkable success in describing the cosmic microwave background (CMB) observations over the past three decades. From \textit{COBE} to \textit{WMAP} and \textit{Planck} satellites, the measured temperature anisotropy power spectrum has consistently converged toward the six-parameter cosmological model \cite{Bennett:1996ce,WMAP:2012fli,aghanimPlanck2018Results2020a}. 

Despite this overall agreement, a collection of large-scale statistical anomalies has persisted across multiple generations of experiments, deviating from the $\Lambda$CDM model or the assumption of an isotropic universe \cite{WMAP:2003ivt,Bennett:2010jb,Planck:2013lks,Planck:2015igc,Planck:2019evm,Herold:2025mro,S1}: a lack of two-point correlations on scales beyond $60^\circ$ \cite{Hinshaw:1996ut,Copi:2008hw,Copi:2013zja}, the alignment between the quadrupole and octopole moments \cite{Tegmark:2003ve,deOliveira-Costa:2003utu,Schwarz:2004gk}, a hemispherical power asymmetry (or dipole modulation) \cite{Hansen:2004vq,Eriksen:2003db,Hoftuft:2009ApJ}, a point-parity asymmetry between odd and even spherical harmonic multipoles \cite{Land:2005jq,Kim:2010gf,Gruppuso:2010nd,P1,P2,P3}, and cold spots \cite{Vielva:2003et,Cruz:2004ce,Cayon:2005er,C1,C2}. Although the statistical significance of these anomalies remains modest ($\sim2\text{--}3\sigma$) \cite{Planck:2019evm} (even lower after correcting for the ``look‑elsewhere'' effect), below the 5$\sigma$ threshold typically required to claim new physics, their consistency across diverse experiments and component separation methods suggests that they are not attributable to instrumental systematics or residual Galactic foregrounds\footnote{It remains possible that some of the anomalies arise from a secondary effect such as the Integrated Sachs-Wolfe effect \cite{Inoue:2006rd,Francis:2009pt,Rassat:2013caa,Naidoo:2017woy}, or nearby galaxies cooling the large-scale CMB temperature \cite{Luparello:2022kqb,Hansen:2023gra,Cruz:2024xbh,Hansen:2024vgs}.}, warranting a careful reassessment of their origin.

A fundamental limitation in interpreting the CMB anomalies is that the CMB temperature on large angular scales has already been measured by \textit{Planck} to near cosmic-variance-limited precision. Consequently, the temperature data alone cannot decisively discriminate between a rare statistical fluctuation---the so-called ``fluke hypothesis''---and a genuine signature of physics beyond the standard model---a thrilling prospect we would welcome. The CMB polarization provides an independent observational avenue to break the impasse. The large-scale $E$-mode signal originates from the same primordial density perturbations as the temperature anisotropies, but sourced by distinct moments, thereby circumventing the \textit{a posteriori} selection effect\footnote{The application of many statistical tests to detect anomalies inevitably yields a small number of $\sim3\sigma$ outliers \cite{Guth:2026qiu}.} in the temperature anomaly analyses. 

The \textit{Planck} 2018 polarization data offer the first opportunity to examine the large-scale anomalies identified in temperature through $E$-mode measurements. However, the sensitivity of these analyses is limited by instrumental noise and residual systematics, particularly on the largest scales where the $E$-mode signal is faintest. As a result, the significance of any test for non-Gaussianity or anomaly remains inconclusive, with results varying across component separation methods. On-going and next-generation CMB polarization experiments, including ACT \cite{Henderson:2015nzj,AtacamaCosmologyTelescope:2025vnj}, Simons Observatory (SO) \cite{SimonsObservatory:2018koc}, CLASS \cite{Harrington:2016jrz,Dahal:2021uig}, AliCPT \cite{liProbingPrimordialGravitational2019}, \textit{LiteBIRD} \cite{Hazumi:2019lys} and etc, are expected to overcome these limitations.

In this paper, we assess the capability of the Ali CMB Polarization Telescope (AliCPT) to constrain large-scale $E$-mode polarization anomalies. As currently the only ground-based CMB experiment located in the Northern Hemisphere, AliCPT enables observations covering $\sim40\%$ of the sky, providing a critical complement in sky coverage to southern-hemisphere experiments such as CLASS and SO. This advantage is particularly valuable for testing anomalies, since the uncertainty of large-scale statistics is primarily limited by the accessible sky fraction.

Given the correlation between the temperature and polarization fields, it is worth noting that the $E$-mode field---whose corresponding temperature field already exhibits known anomalies---may behave differently from a realization that fluctuates freely within the $\Lambda$CDM model. We refer to the former as ``constrained'' realizations \cite{Copi:2013zja,Chiocchetta:2020ylv}, in which part of the polarization field is constrained by the observed temperature anisotropies, and the latter as ``unconstrained'' realizations.

Recent studies \cite{Shi:2022hxc,LiteBIRD:2025tnn} have shown that the constrained $E$-mode realizations, conditioned on the \textit{Planck} SMICA temperature data, yield distributions of the anomaly statistics that are statistically indistinguishable from those obtained from unconstrained simulations. Only cross-$TE$ estimators may distinguish between the two at a moderate level of significance \cite{LiteBIRD:2025tnn}. In other words, under the fluke hypothesis, the $E$ modes constrained by current temperature observations are not expected to exhibit statistical excursions that deviate significantly from $\Lambda$CDM predictions.

This finding simplifies the forecasting task, as our objective is to evaluate whether AliCPT can reject the fluke hypothesis based on anomaly statistics derived from $E$-mode polarization data. Constrained realizations, while physically motivated, do not significantly alter the statistical distributions of interest, and are therefore not required for the sensitivity forecasts presented here. Accordingly, we apply a suite of statistical tests to unconstrained CMB simulations for AliCPT (considering two noise levels, baseline and goal), \revise{SO (goal) and the joint AliCPT+SO dataset}. For each test, we compute the statistical distribution of $\Lambda$CDM simulations; the anomaly statistic derived from actual $E$-mode data will be compared to this distribution to determine whether the fluke hypothesis is rejected.

The remainder of this paper is organized as follows. In Section~\ref{sec:sim-msk}, we describe the simulations and masks used in this work, detailing the generation of component-separated maps and their corresponding power spectra. Section~\ref{sec:est-ano} introduces the estimators for the large-scale anomalies considered: dipole modulation, lack of large-angle correlations, alignment of the quadrupole and octopole, and point-parity asymmetry. In Section~\ref{sec:rest}, we present our results, including validation of the local variance estimator using dipole-modulated realizations, and a comparison of the statistical distributions derived from unconstrained simulations at different sensitivity levels. We conclude in Section~\ref{sec:concl} with a summary of our main findings.

\section{Simulations and masks}
\label{sec:sim-msk}

\subsection{Simulations}

\begin{figure}[tbp]
    \centering
    \begin{subfigure}[b]{0.49\textwidth}
        \centering
        \includegraphics[width=\textwidth]{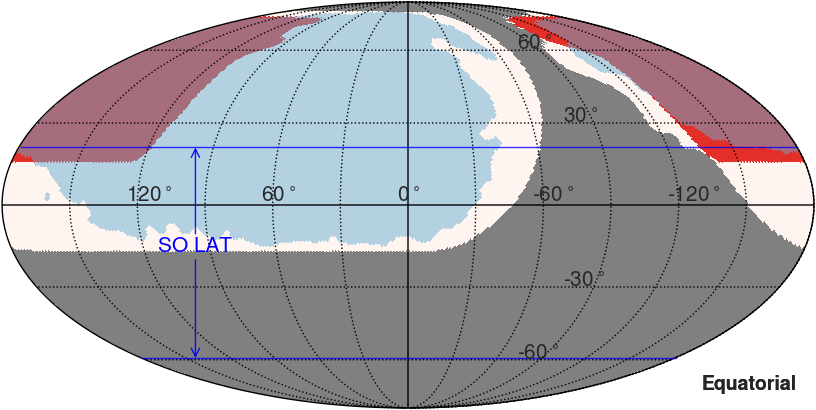}
    \end{subfigure}
    \begin{subfigure}[b]{0.49\textwidth}
        \centering
        \includegraphics[width=\textwidth]{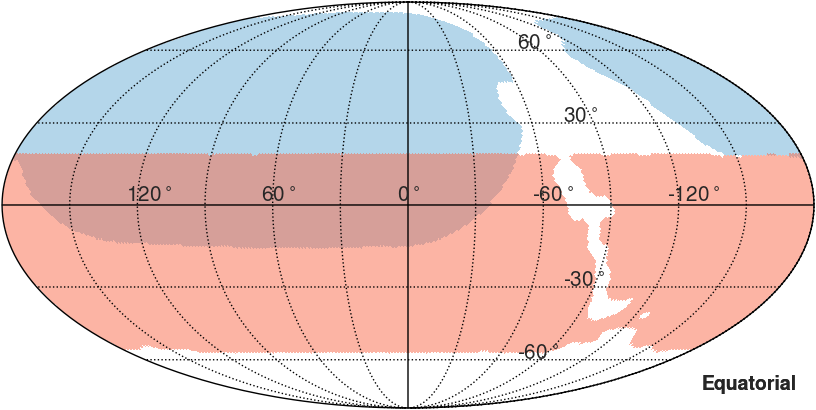}
    \end{subfigure}
    \caption{\textit{Left}: AliCPT footprints in Equatorial coordinates. The white and red areas denote the wide-scan (WS) and deep-scan (DS) coverage, respectively. The light blue region corresponds to the confidence mask applied in the WS baseline $E$-mode analyses. The SO LAT survey region is indicated between blue horizontal lines. \textit{Right}: AliCPT WS (light blue) and SO (light red) respective confidence masks for goal $E$-mode analyses.}
    \label{fig:msk}
\end{figure}

\begin{table}[tbp]
    \centering
    \caption{Instrumental characteristics of the \textit{WMAP}, AliCPT WS, SO LAT, and \textit{Planck} HFI frequency channels, with $\sigma_n^{\rm P}$ denoting the polarization noise level, which is $\sqrt{2}$ times the temperature noise level for AliCPT and SO. The baseline and goal noise cases for AliCPT correspond to the first-year and four-year AliCPT observations, respectively, representing an integrated sensitivity of 4 and 48 mod$\cdot$yr. The SO LAT sensitivities are taken from Ref.~\cite{SimonsObservatory:2025wwn}.}
    \begin{tabular}{l|c|c|c|c}
        \hline
        \hline
        \makecell{Experiment} & 
        \makecell{Frequency \\ {[GHz]}} & 
        \makecell{FWHM \\ {[arcmin]}} & 
        \makecell{$\sigma_n^{\rm P}$ (baseline) \\ {[$\mu$K-arcmin]}} & 
        \makecell{$\sigma_n^{\rm P}$ (goal) \\ {[$\mu$K-arcmin]}} \\
        \hline
        WMAP   & 23  & 52.8 & 496 & -  \\
        \hline
        \multirow{2}{*}{AliCPT WS} & 95  & 19   & 34  & 10 \\
                                 & 150 & 11   & 44  & 13 \\
        \hline
        \multirow{6}{*}{SO LAT}      & 27   & 7.4    & 86   & 62  \\
                                 & 39   & 5.1    & 42   & 33  \\
                                 & 93   & 2.2    & 7.5  & 5.4  \\
                                 & 145  & 1.4    & 9.3  & 5.8  \\
                                 & 225  & 1.0    & 21   & 14  \\
                                 & 280  & 0.9    & 49   & 35  \\
        \hline
        \multirow{4}{*}{Planck HFI} & 100 & 9.7 & 81  & - \\
                                    & 143 & 7.3 & 66  & - \\
                                    & 217 & 5.0 & 92  & - \\
                                    & 353 & 4.9 & 399 & - \\
        \hline
    \end{tabular}
    \label{tab:instr}
\end{table}

The AliCPT footprint covers a sky fraction of 52\% including most of the northern sky, referred to as the ``wide-scan'' (WS) patch, as shown in Fig.~\ref{fig:msk}. Besides, AliCPT will deep-scan a 13\% sky patch with the cleanest foreground contamination among the northern sky and a lower noise level compared to the wide scan, mainly for the purpose of detecting primordial $B$ modes \cite{liProbingPrimordialGravitational2019,Dou:2024spy}. However, the deep-scan scenario is not suitable for measuring largest-scale $E$ modes due to the limited sky coverage. Therefore, in this work, we focus on the WS coverage.

We consider two noise cases for AliCPT: the ``baseline'' and ``goal'' scenarios, corresponding to the sensitivities of the first-year and four-year WS surveys, respectively. Their noise levels are listed in Table~\ref{tab:instr}. AliCPT's first-year observations, based on four detector modules, provide a sensitivity of 4 mod$\cdot$yr. With the annual increment of detector modules over the four-year survey, the total cumulative sensitivity of the goal case amounts to 48 mod$\cdot$yr. 

To remove foreground contamination, similar to previous works \cite{Dou1,Dou2,Dou3}, we combine AliCPT simulations with two external datasets (using NILC detailed in Section~\ref{sec:nilc}): the 23 GHz K band of \textit{WMAP}, and the four High Frequency Instrument (HFI) bands at 100, 143, 217, and 353 GHz from the \textit{Planck} Public Release 4 (PR4) \cite{Planck:2020olo}. Their instrumental parameters are summarized in Table~\ref{tab:instr}.

We simulate the Galactic foregrounds using the \texttt{PySM}\footnote{\url{https://pysm3.readthedocs.io/en/latest/}} \cite{thornePythonSkyModel2017,zoncaPythonSkyModel2021,groupFullskyModelsGalactic2025} \texttt{d9s4f1a1co1} model, which includes thermal dust, synchrotron, free-free, anomalous microwave emission (AME), and CO line emissions. For the polarized components, the model adopts the \textit{Planck} GNILC thermal dust template scaled by a modified blackbody spectrum, and the \textit{WMAP} 9-year synchrotron template scaled by a power-law SED with fixed spectral indices\footnote{Specifically, $\beta_d=1.48$, $T_d=19.6\,\mu$K for dust, and $\beta_s=-3.1$ for synchrotron.}. The free-free, AME, and CO line emissions are modeled as unpolarized components.

We generate 1000 Gaussian CMB realizations based on the power spectra derived from the \textit{Planck} best-fit $\Lambda$CDM model\footnote{The six $\rm \Lambda$CDM parameters are: dark matter density $\Omega_ch^2=0.120$, baryon density $\Omega_bh^2=0.02237$, scalar spectral index $n_s=0.9649$, optical depth $\tau=0.0544$, Hubble constant $H_0=67.36\,{\rm km\,s^{-1}\,Mpc^{-1}}$ and the primordial comoving curvature power spectrum amplitude $A_s=2.10\times10^{-9}$.} \cite{aghanimPlanck2018Results2020a}. In Sect.~\ref{sec:dm-val}, we generate a separate dataset of modulated CMB realizations, designed to validate the dipole modulation estimator. For the noise simulations, we produce 100 Gaussian noise realizations
using the hitmaps of AliCPT and \textit{WMAP}, and directly adopt the PR4 NPIPE noise simulations provided on the \textit{Planck} Legacy Archive (PLA)\footnote{\url{https://pla.esac.esa.int}}. Finally, the noise realizations are randomly combined with the 1000 CMB simulations and added to the foregrounds, yielding 1000 simulated sky maps (with two datasets for AliCPT corresponding to two noise cases). All of the above maps are converted to a \texttt{HEALPix}\footnote{\url{https://healpix.sourceforge.io/}} \citep{gorskiHEALPixFrameworkHigh2005} resolution of $N_{\text{side}} = 1024$ to match the AliCPT's original resolution. They will be downgraded to $N_{\text{side}} = 64$ resolution after component separation.

We additionally apply our statistical tests to \textit{Planck} SMICA simulations using the \textit{Planck} common mask, serving as a sensitivity comparison with AliCPT. The 300 PR3 SMICA noise simulations obtained from PLA are randomly combined with 1000 CMB realizations.

\subsection{NILC foreground cleaning}
\label{sec:nilc}

\begin{figure}[tbp]
    \centering
    \begin{subfigure}[b]{0.47\textwidth}
        \centering
        \includegraphics[width=\textwidth]{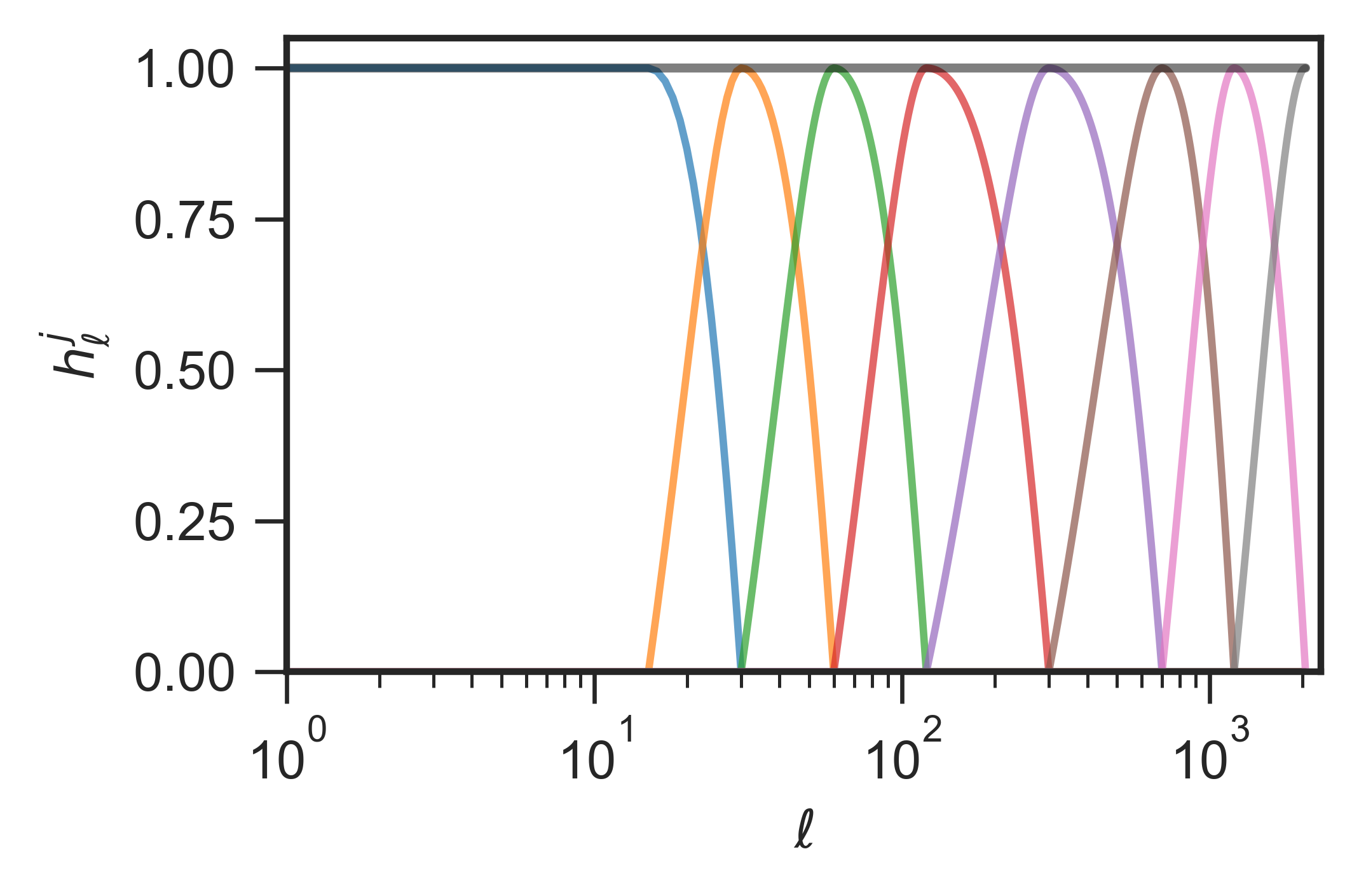}
    \end{subfigure}
    \begin{subfigure}[b]{0.51\textwidth}
        \centering
        \includegraphics[width=\textwidth]{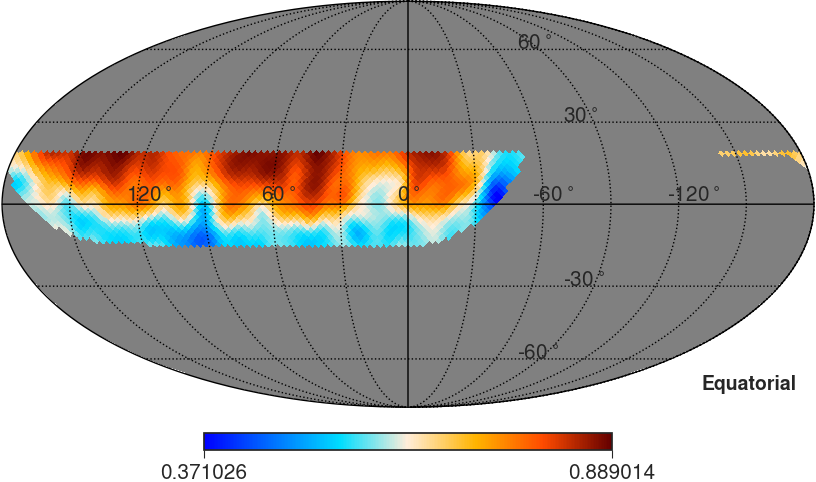}
    \end{subfigure}
    \caption{\textit{Left}: Cosine needlet bands $h_\ell^j$ adopted in the NILC pipeline. \textit{Right}: The inverse variance weights of AliCPT at the first needlet band. The weights of AliCPT and SO sum to unity.}
    \label{fig:nilc}
\end{figure}

We implement the needlet internal linear combination (NILC) \cite{delabrouilleFullSkyLow2009,basakNeedletILCAnalysis2012,basakNeedletILCAnalysis2013} technique to remove diffuse Galactic foregrounds for both $T$ and $E$ maps. First, we apply a mask discarding pixels with a polarization noise standard deviation greater than 60 $\mu$K at the 150 GHz band.

The input multi-frequency maps, originally at different angular resolutions, are then convolved to a common beam using $d_{\ell m}^{\nu,\rm out}=d_{\ell m}^{\nu,\rm in}b_\ell^{\rm out}/b_\ell^{\rm in}$, where $b_\ell$ denotes the beam transfer function and $d_{\ell m}^{\nu}$ the spherical harmonic coefficients for $\nu$ channel. The output resolution ($b_\ell^{\rm out}$) is set to a Gaussian beam with FWHM of 11$'$, matching the 150 GHz channel. For the \textit{WMAP} K and AliCPT 95 GHz bands with larger beam sizes (i.e., $b_\ell^{\rm in}<b_\ell^{\rm out}$), we truncate the harmonic coefficients at $\ell>350$ and $\ell>1200$, respectively, to prevent excessive noise amplification.

To achieve harmonic-space localization, we decompose the spherical harmonic coefficients into a set of filtered maps via:
\begin{equation}\label{eq:dlm_nu}
    d_{\ell m}^{\nu,j}=h_\ell^j d_{\ell m}^\nu\,,
\end{equation}
where the needlet bands $h_\ell^j$ form a partition of unity, i.e., $\sum_j(h_\ell^j)^2=1$. We employ cosine needlet bands with peaks located at multipoles 15, 30, 60, 120, 300, 700, 1200, and $\ell_{\max}=2048$, as shown in Fig.~\ref{fig:nilc}. For the \textit{WMAP} K band, where multipoles above 350 have been removed, needlets with peak multipoles exceeding 350 are excluded from the NILC computation (similarly for the 95 GHz channel). After filtering, $d_{\ell m}^{\nu,j}$ is transformed back to the pixel domain, yielding a set of needlet maps.

Once in needlet space, we perform an internal linear combination across frequencies. For each needlet scale $j$ and pixel $k$, we compute the empirical frequency–frequency covariance matrix $\hat{\boldsymbol C}_{jk}$ by averaging the product of needlet maps of two frequencies over a circular patch centered on pixel $k$. The needlet maps are then combined using NILC weights:
\begin{equation}
    w^{\rm NILC}_{\nu, j}(\hat{\boldsymbol k}) = \left[\frac{\hat{\boldsymbol C}_{jk}^{-1} \boldsymbol a}{\boldsymbol a^t\hat{\boldsymbol C}_{jk}^{-1} \boldsymbol a}\right]_\nu\,,
    \label{eq:nilc_wgts}
\end{equation}
where $\boldsymbol a=[1,\dots,1]^t$ is the spectral response vector encoding the frequency dependence of the desired signal, and $^t$ denotes the transpose. The constraint $\sum_\nu a_\nu w^{\rm NILC}_{\nu, j}=1$ guarantees that the CMB component is recovered with unit gain.

Finally, an inverse needlet transform in harmonic space reconstructs the NILC-cleaned map. The cleaned map, originally at 11$'$ resolution with $N_{\text{side}}=1024$, is subsequently degraded to a lower resolution of 160$'$ with $N_{\text{side}}=64$ for real space analyses:
\begin{equation}
    {\hat s}_{\ell m}^{\rm NILC,out}={\hat s}_{\ell m}^{\rm NILC,in}\frac{b_\ell^{\rm out}p_\ell^{\rm out}}{b_\ell^{\rm in}p_\ell^{\rm in}}\,,
\end{equation}
where $p_\ell$ denotes the \texttt{HEALPix} pixel window function.

\subsection{Confidence masks}

\begin{table}[tbp]
    \centering
    \caption{Sky fractions [\%] of $E$-mode confidence masks for AliCPT baseline and goal noise schemes, \revise{the SO-goal configuration}, and the combination of AliCPT and SO goal sensitivities. The values in parentheses represent sky fractions of the wide-scan footprints \revise{(without foreground masking)} for AliCPT, SO, and AliCPT+SO.}
    \begin{tabular}{c| c c c c}
        \hline
        \hline
        $N_{\rm side}$ & Ali-baseline & Ali-goal & SO-goal & Ali+SO \\
        \hline
        $64$ & 38 (43) & 42 (43) & 56 (59) & 82 (85) \\
        \hline
        $32$ & 35 (40) & 39 (40) & 51 (57) & 78 (83) \\
        \hline
    \end{tabular}
    \label{tab:fsky}
\end{table}

After component separation, we construct confidence masks by thresholding the foreground-plus-noise residual RMS maps derived from simulations, followed by smoothing and binary thresholding to remove isolated regions. This procedure is implemented separately for each noise scheme (baseline and goal), yielding masks with distinct sky fractions. Further details are provided in Appendix~\ref{app:con-msk}, the resulting confidence masks are shown in Fig.~\ref{fig:msk}, and the sky fractions are summarized in Table~\ref{tab:fsky}. Actually, confidence masks are applied only in the dipole modulation analysis, since residual foreground contamination is found to be negligible for both the large-scale power spectra at $\ell\lesssim30$ (\revise{i.e., the bias is negligible compared to the error bar, see the bottom panels of Fig.~\ref{fig:dl-xqml}}) and the quadrupole-octopole alignment statistic.

\revise{We only consider a low-complexity foreground model here, without including extragalactic point sources or various instrumental systematics. We defer the consideration of residual foreground spectra arising from more complex foreground models to future work. However, given that NILC is a blind component separation method that does not rely on specific foreground models, we expect that real foregrounds, after applying a Galactic mask of a few percent, will not introduce a significant bias in the $E$-mode power spectra for AliCPT or SO.}

\subsection{Combination with SO LAT}
To supplement the missing information in the southern sky, we incorporate the Simons Observatory (SO) Large Aperture Telescope (LAT) \cite{SimonsObservatory:2018koc,SimonsObservatory:2025wwn} simulations, with instrumental parameters presented in Table~\ref{tab:instr}. In this work we only focus on the ``goal'' sensitivity of SO. The SO survey covers 61\% of the sky with DEC ranging from $-59^\circ$ to $21^\circ$, overlapping with the AliCPT footprint by 24\% of the sky. The original LAT maps are provided at $N_{\text{side}}=2048$ for 27 and 39 GHz bands, and $N_{\text{side}}=8192$ for the rest. For computational efficiency, we directly generate low-resolution LAT simulations at $N_{\text{side}}=64$ and 160$'$ FWHM\footnote{The bandpass weights of SO LAT are obtained from \url{https://github.com/simonsobs/map_based_simulations/tree/main/mbs-s0012-20230321/simonsobs_instrument_parameters_2023.03}.}. The CMB and foreground realizations are produced using the same pipeline as AliCPT, while the noise power spectrum is given by:
\begin{equation}
    N_\ell=N_{\text{red}}\left(\frac{\ell}{\ell_{\text{knee}}}\right)^{\alpha_{\text{knee}}}+N_{\text{white}}\,,
\end{equation}
where $N_{\text{white}}$ denotes the white noise component with noise levels listed in Table~\ref{tab:instr}, and $N_{\text{red}}$, $\ell_{\text{knee}}$, and $\alpha_{\text{knee}}$ characterize the $1/f$ noise originating from atmospheric and electronic noise, with values taken from Ref.~\cite{SimonsObservatory:2018koc}: $N_{\text{red}}=N_{\text{white}}$, $\ell_{\text{knee}}=700$, and $\alpha_{\text{knee}}=-1.4$ for polarization. We do not model $1/f$ noise for temperature, since the large-scale CMB temperature is already well measured by \textit{WMAP} and \textit{Planck}. 

Similarly to AliCPT, we combine SO with \textit{Planck} HFI simulations using NILC\footnote{Only the first four needlet bands in Fig.~\ref{fig:nilc} are employed here, as the input resolution is $N_{\text{side}}=64$.} to suppress the large-scale noise. The SO confidence mask shown in Fig.~\ref{fig:msk} is determined by thresholding the RMS residual map at 0.24 $\mu$K for $E$ modes. Finally, We combine AliCPT and SO NILC-cleaned maps (with goal sensitivities) using the following strategies: in non-overlapping regions between AliCPT and SO, the respective cleaned map are directly adopted; for overlapping regions, we implement inverse-noise-variance-weighted combinations in the needlet space, with the weights of SO at $j$ needlet scale given by:
\begin{equation}
    w_{j,{\rm SO}}(\hat{\boldsymbol k}) = \frac{\left[\sigma^{2}_{j,{\rm SO}}(\hat{\boldsymbol k})\right]^{-1}}{\left[\sigma^{2}_{j,{\rm SO}}(\hat{\boldsymbol k})\right]^{-1}+\left[\sigma^{2}_{j,{\rm Ali}}(\hat{\boldsymbol k})\right]^{-1}}\,,
\end{equation}
where the variance at $j$ needlet scale and $k$ pixel, $\sigma^{2}_{j}(\hat{\boldsymbol k})$, is derived from the foreground-plus-noise residual of simulations after NILC cleaning. For instance, the first-needlet-band weights of AliCPT are shown in Fig.~\ref{fig:nilc}.

\revise{In addition to the joint analysis, we also treat the SO‑goal configuration as an independent experiment. For this standalone case, we use the same NILC‑cleaned SO maps and apply the SO confidence mask described above. This enables a direct comparison between SO-only and the combined AliCPT+SO dataset, allowing us to quantify the added detection power of AliCPT.}

\subsection{xQML power spectra}

\begin{figure}[tbp]
    \centering
    \begin{subfigure}[b]{0.49\textwidth}
        \centering
        \includegraphics[width=\textwidth]{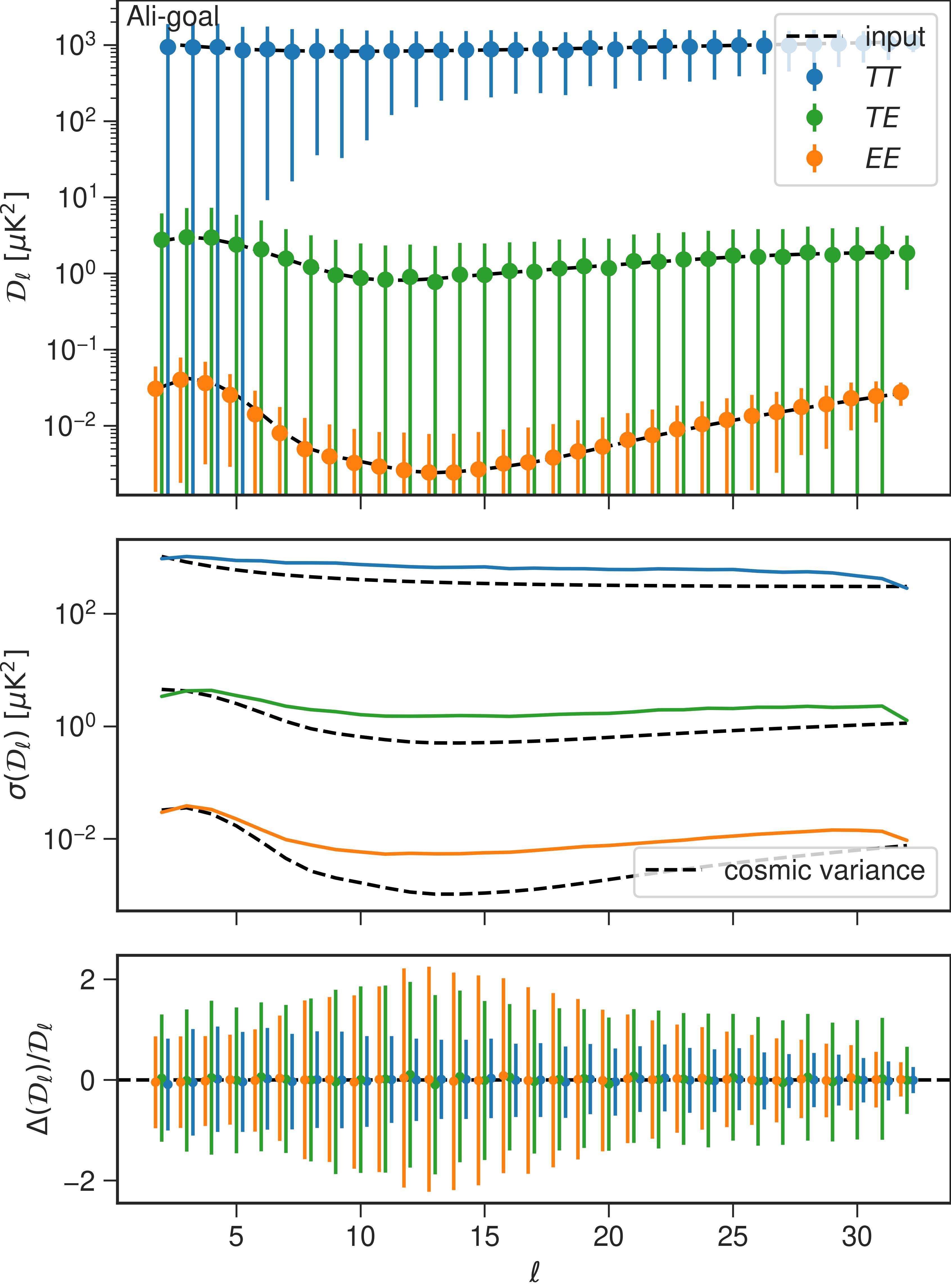}
    \end{subfigure}
    \begin{subfigure}[b]{0.49\textwidth}
        \centering
        \includegraphics[width=\textwidth]{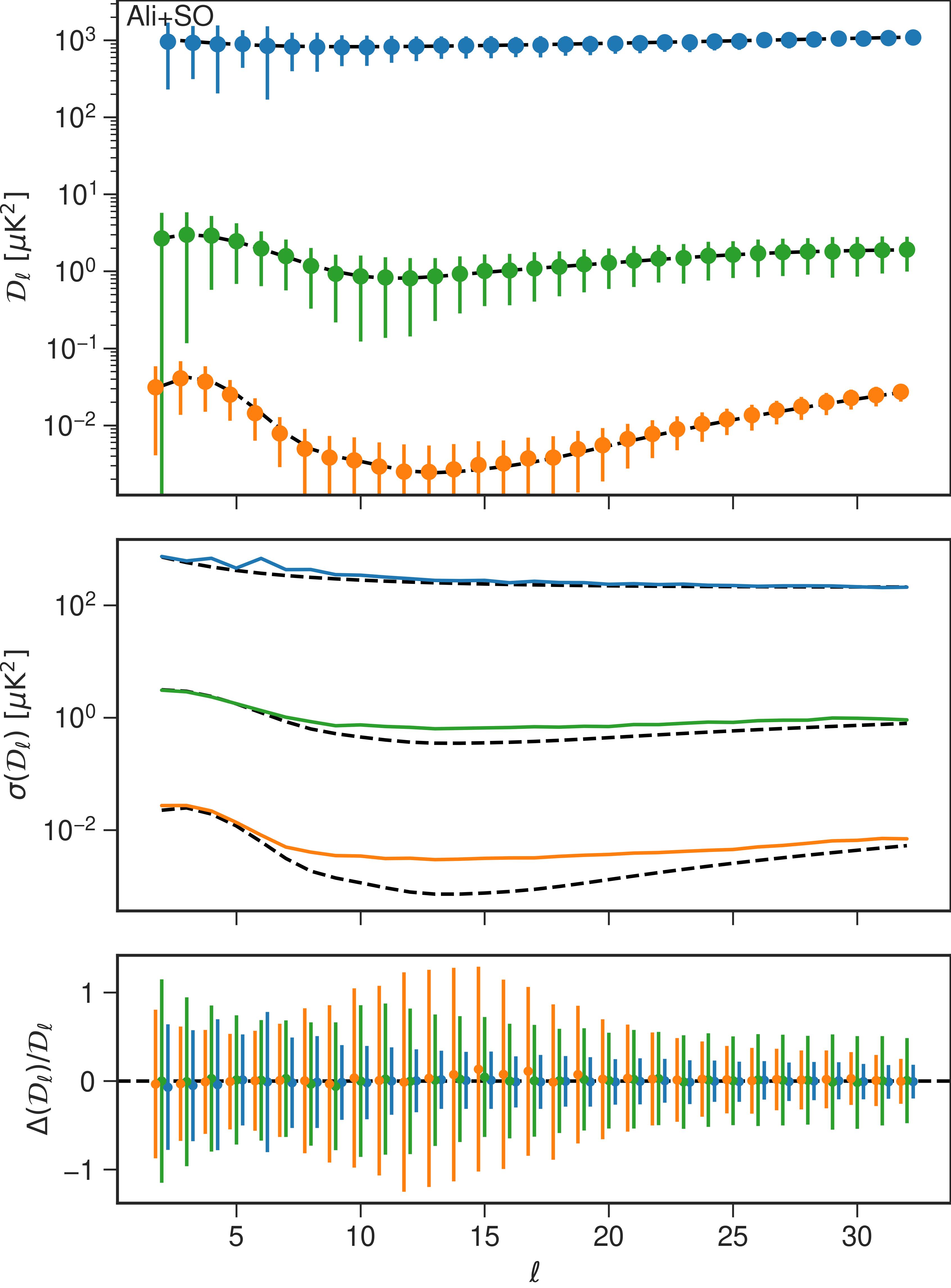}
    \end{subfigure}
    \caption{\textit{Top}: \texttt{xQML} power spectra computed from NILC-cleaned realizations for goal-AliCPT (left) and AliCPT+SO (right) noise schemes. \revise{\textit{Middle}: Standard deviations of the power spectra compared to the cosmic variance (black dashed curve). \textit{Bottom}: Relative bias of power spectra (i.e., bias divided by the theoretical CMB power spectra), where the error bars represent the relative standard deviation $\sigma(\mathcal{D}_\ell)/\mathcal{D}_\ell$.} The large-scale power spectra for baseline-AliCPT are similar to those for goal-AliCPT.}
    \label{fig:dl-xqml}
\end{figure}

For the harmonic-based anomaly estimators (i.e., large-angle correlations and point-parity asymmetry), we compute the CMB power spectra using \texttt{xQML}\footnote{\url{https://gitlab.in2p3.fr/xQML/xQML}} \cite{Vanneste:2018azc}, a cross-spectrum quadratic maximum-likelihood (QML) estimator. A key strength of the QML method over the pseudo-$C_\ell$ (PCL) approach is its optimal uncertainty on large angular scales \cite{Tegmark2001}. However, the QML estimator requires $\mathcal{O}(N_{\rm pix}^3)$ operations for a map of pixel number $N_{\rm pix}$, making it more computationally expensive than PCL and only feasible for low-resolution analyses \cite{Tegmark2001,Chen1,Chen2}. Therefore, the NILC maps are downgraded to \revise{$N_{\text{side}}=32$ and smoothed with a 320$'$ Gaussian beam}, from which the $EE$ and $TE$ power spectra are calculated at $\ell\in[2, 32]$. \revise{Appendix~\ref{app:just-repix} demonstrates that this repixelization rarely alters estimated power spectra.}

The QML estimator requires precise knowledge of the noise pixel covariance matrix $\mathbf{N}$ to remove the noise bias. Obtaining this is challenging due to the instrumental complexity. The \texttt{xQML} method, however, cross-correlates two data split maps with the same signal but uncorrelated noise components to null the noise bias, rendering the estimator unbiased independent of the choice of $\mathbf{N}$. Throughout this work, we multiply the full-mission noise simulations by a factor of $\sqrt{2}$ to mimic A/B-split noise simulations, and approximate the noise covariance matrix $\mathbf{N}$ of A/B splits by its diagonal part computed from 100 NILC noise simulations. Finally, we add the same CMB realization to two different noise splits and use the \texttt{xQML} method to estimate their cross $EE$ power spectra. We still use full-mission maps to compute the $TE$ power spectra, as the noise bias is automatically canceled.

We apply the wide-scan masks (see Table~\ref{tab:fsky} for sky fractions) instead of the confidence masks to NILC maps before computing the power spectra, since we have found that the residual foreground bias on the polarization power spectra is negligible (see \revise{the bottom panels of} Fig.~\ref{fig:dl-xqml}). We also observe that the goal results are similar to the baseline ones for AliCPT, because at the largest scales, the uncertainties of power spectra are dominated by sky fraction (or cosmic variance) rather than noise. In Fig.~\ref{fig:dl-xqml}, the $TT$ power spectra are shown for illustration only, as they are not used in our analysis and $1/f$ noise is omitted. In addition, due to the limitation of the NILC method on a partial sky, the AliCPT(+SO) $T$ maps still have higher large-scale noise compared to \textit{Planck}. Therefore, for the computation of $TE$ power spectra, we employ cross-spectra between the \textit{Planck} SMICA temperature simulations and the AliCPT(+SO) $E$-mode simulations, applying both the \textit{Planck} inpainting mask and the wide-scan mask in order to maximize the sky fraction.

\section{Estimators of CMB anomalies}
\label{sec:est-ano}

\subsection{Dipole modulation}
\label{sec:dm}

Dipole modulation, or hemispherical power asymmetry, was first detected in the first-year \textit{WMAP} temperature data \cite{Hansen:2004vq,Eriksen:2003db}. Specifically, the power spectrum (or variance in the real space) estimated from one hemisphere was found to be larger than that from the opposite hemisphere over the multipole range $\ell=2-40$. A phenomenological model was proposed to resolve the anomaly \cite{Gordon:2005ai}:
\begin{equation}\label{eq:dm}
    T^{\rm obs}(\hat{\boldsymbol n})=\tilde T_{\rm CMB}(\hat{\boldsymbol n})[1+\boldsymbol d\cdot\hat{\boldsymbol n}]+T_N(\hat{\boldsymbol n})\,,
\end{equation}
where $T^{\rm obs}(\hat{\boldsymbol n})$ denotes the observed CMB temperature map, $\tilde T_{\rm CMB}(\hat{\boldsymbol n})$ denotes the isotropic CMB signal, $\boldsymbol d$ denotes the modulating dipole, and $T_N(\hat{\boldsymbol n})$ denotes the instrumental noise. The dipole modulation was later reinforced in the \textit{WMAP} \cite{Hoftuft:2009ApJ,Hanson:2009PRD} and \textit{Planck} \cite{Akrami:2014eta,Planck:2019evm} temperature data, revealing a modulating amplitude of about 7\% in a direction around $(209^\circ, -15^\circ)$ with a significance approaching $3\sigma$. 

For \textit{Planck} $E$-mode polarization, however, the $p$-values from the four component-separation methods present notable differences, which argues against a robust detection of cosmological power asymmetry in polarization. Nevertheless, the dipole directions derived from the $E$-mode maps were found to be remarkably close to those from the temperature data, suggesting a potential alignment that warrants further scrutiny \cite{Planck:2019evm,Gimeno-Amo:2023jgv}. Ref.~\cite{Gimeno-Amo:2025icf} reports an anomalous dipole in the amplitude of the primordial power spectrum, $A_s$, whose direction aligns with that of the hemispherical power asymmetry.

In our analysis, we assume the polarization fields follow the same dipole modulation pattern as the temperature
in Eq.~\eqref{eq:dm} \cite{Ghosh:2015qta,Ghosh:2018apx}:
\begin{equation}
    P^{\rm obs}_{\pm}(\hat{\boldsymbol n})=\tilde P_{\pm,\rm CMB}(\hat{\boldsymbol n})[1+\boldsymbol d\cdot\hat{\boldsymbol n}]+P_{\pm,N}(\hat{\boldsymbol n})\,,
\end{equation}
where $P_{\pm}(\hat{\boldsymbol n})=Q(\hat{\boldsymbol n})\pm iU(\hat{\boldsymbol n})$.
We adopt a pixel-based method called the local variance estimator (LVE) \cite{Akrami:2014eta} to measure the power asymmetry. First, we remove the monopole and dipole from the observed map after masking. The $E$-mode local variance map (LVM) is then defined as\footnote{Here the approximation holds when $\|{\boldsymbol d}\|\ll1$ and the $B$-mode field is negligible compared to the $E$-mode field.}:
\begin{equation}\label{eq:lvm}
    \sigma^2_r(\hat{\boldsymbol n})\equiv\frac{1}{N_{k}}\sum_{\hat{\boldsymbol k}\in\bigodot_{r,\hat{\boldsymbol n}}}\big[E^{\rm obs}(\hat{\boldsymbol k})-\bar{E}\big]^2\approx\tilde \sigma_{r,\rm CMB}^2(\hat{\boldsymbol n})[1+2\boldsymbol d\cdot\hat{\boldsymbol n}]+\sigma^2_{r,N}(\hat{\boldsymbol n})\,,
\end{equation}
where $\bigodot_{r,\hat{\boldsymbol n}}$ includes all pixels (of number $N_k$) within an $r$-radius circular area centered at $\hat{\boldsymbol n}$, $\bar{E}$ denotes the observed $E$-mode field averaged over the pixels inside the disc region, $\tilde \sigma_{r,\rm CMB}^2(\hat{\boldsymbol n})$ and $\sigma^2_{r,N}(\hat{\boldsymbol n})$ denote the local variance maps of the isotropic CMB and noise, respectively. In practice, we only compute the variance for the circular discs within which no more than 90\% of the pixels are masked. We note that the \texttt{HEALPix} resolutions associated with $\hat{\boldsymbol n}$ and $\hat{\boldsymbol k}$ may differ: $\hat{\boldsymbol n}$ is defined as the pixel centroids of a lower-resolution \texttt{HEALPix} map compared to that of the original map $E^{\rm obs}(\hat{\boldsymbol k})$. 
We also compute the average and variance of the LVMs derived from isotropic simulations, represented by $\langle\sigma^2_{r,{\rm iso}}(\hat{\boldsymbol n})\rangle$ and $\sigma^2\big(\sigma^2_{r,{\rm iso}}(\hat{\boldsymbol n})\big)$, respectively. The LVMs of isotropic simulations are given by:
\begin{equation}\label{eq:lvm-iso}
    \sigma^2_{r,{\rm iso}}(\hat{\boldsymbol n})=\tilde \sigma_{r,\rm CMB}^2(\hat{\boldsymbol n})+\sigma^2_{r,N}(\hat{\boldsymbol n})\,.
\end{equation}
Although the anisotropic noise and Doppler boosting are incorporated in simulations, we refer to these simulations as ``isotropic'' to emphasize the isotropy of CMB realizations. As a contrast, we generate ``modulated'' simulations with an assumed dipole injection that will be discussed in Section~\ref{sec:dm-val}. Using Eqs.~\eqref{eq:lvm} and \eqref{eq:lvm-iso}, the un-normalized dipole modulation is estimated by fitting a dipole to the local variance map with inverse variance weights after correcting for the noise bias:
\begin{equation}\label{eq:dm-unnorm}
    \boldsymbol p = \mathop{\arg\min}\limits_{\boldsymbol p}\bigg\{\sum_{\hat{\boldsymbol n}}\frac{\big[\sigma^2_r(\hat{\boldsymbol n})-\langle\sigma^2_{r,{\rm iso}}(\hat{\boldsymbol n})\rangle-p_0-\boldsymbol p\cdot\hat{\boldsymbol n}\big]^2}{\sigma^2\big(\sigma^2_{r,{\rm iso}}(\hat{\boldsymbol n})\big)}\bigg\}\,.
\end{equation}
This step is implemented by the \texttt{remove\_dipole} routine of the \texttt{HEALPix} software. We note that this dipole $\boldsymbol p$ is expected to have the same direction but different amplitude than the dipole modulation $\boldsymbol d$, i.e., $\boldsymbol p =2\tilde \sigma_{r,\rm CMB}^2\boldsymbol d$, assuming $\tilde \sigma_{r,\rm CMB}^2(\hat{\boldsymbol n})$ is isotropic. 

To measure the amplitude of the dipole modulation, we define the normalized local variance map as:
\begin{equation}
    \xi(\hat{\boldsymbol n})\equiv\frac{\sigma^2_r(\hat{\boldsymbol n})-\langle\sigma^2_{r,{\rm iso}}(\hat{\boldsymbol n})\rangle}{\langle\tilde \sigma^2_{r,{\rm CMB}}(\hat{\boldsymbol n})\rangle}\approx 2\boldsymbol d\cdot\hat{\boldsymbol n}\,,
\end{equation}
where $\langle\tilde \sigma^2_{r,{\rm CMB}}(\hat{\boldsymbol n})\rangle$ denotes the average of LVMs obtained from CMB-only realizations based on the best-fit $\Lambda$CDM model. We then fit its dipole with inverse variance weights:
\begin{equation}\label{eq:dm-norm}
    {\boldsymbol d}^{\text{norm}} = \mathop{\arg\min}\limits_{\boldsymbol d}\bigg\{\sum_{\hat{\boldsymbol n}}\frac{\big[\xi(\hat{\boldsymbol n})-d_0-2\boldsymbol d\cdot\hat{\boldsymbol n}\big]^2}{\sigma^2\big(\sigma^2_{r,{\rm iso}}(\hat{\boldsymbol n})\big)/\langle\tilde \sigma^2_{r,{\rm CMB}}(\hat{\boldsymbol n})\rangle^2}\bigg\}\,.
\end{equation}
This normalized dipole can unbiasedly estimate the dipole modulation, i.e., $\mathbb{E}\big[{\boldsymbol d}^{\text{norm}}\big]=\boldsymbol d$. Assuming its $x,y,z$ components $d_{x,y,z}^{\text{norm}}$ are Gaussian variables, the expectation of the squared amplitude of ${\boldsymbol d}^{\text{norm}}$ is given by:
\begin{equation}
    \mathbb{E}\big[\|{\boldsymbol d}^{\text{norm}}\|^2\big]=\|{\boldsymbol d}\|^2+\sigma^2(d_x^{\text{norm}}+d_y^{\text{norm}}+d_z^{\text{norm}})=A_d^2+\mathbb{E}\big[\|{\boldsymbol d}^{\text{norm}}\|^2_{\rm iso}\big]\,,
\end{equation}
where $\|{\boldsymbol d}^{\text{norm}}\|_{\rm iso}$ represents the normalized dipole amplitude obtained from isotropic realizations, and $A_d\equiv\|{\boldsymbol d}\|$ is the real dipole amplitude. Therefore, we correct for the isotropic variance term:
\begin{equation}
    \hat{A}_d=\sqrt{\|{\boldsymbol d}^{\text{norm}}\|^2-\langle \|{\boldsymbol d}^{\text{norm}}\|^2_{\rm iso}\rangle}\,,
\end{equation}
such that $\hat{A}_d^2$ is an unbiased estimate of $A_d^2$, i.e., $\mathbb{E}\big[\hat{A}_d^2\big]=A_d^2$.

Ref.~\cite{LiteBIRD:2025tnn} introduces the results of the local $TE$ covariance maps. We refrain from presenting the local $TE$ covariance here, because $\tilde \sigma_{r,\rm CMB}^{2,TE}(\hat{\boldsymbol n})$ fluctuates around zero across the sky, preventing a meaningful normalization of the dipole.

\subsection{Lack of large-angle correlations}
\label{sec:lolc}

The CMB temperature two-point angular correlation function exhibits a significant deficit of large-angle correlations compared to standard $\Lambda$CDM predictions. This anomaly was first detected in the \textit{COBE-DMR} data \cite{Hinshaw:1996ut,Bennett:1996ce} and subsequently confirmed by \textit{WMAP} \cite{WMAP:2003ivt,Copi:2008hw} and \textit{Planck} \cite{Copi:2013cya,Planck:2019evm}, with $p$-values $\sim1\%$.

Following the definition from the WMAP team \cite{WMAP:2003elm,Copi:2013zja}, we employ the $S^{XY}$ statistic integrating the squared two-point correlations:
\begin{equation}\label{eq:s12}
    S^{XY}(\theta_1,\theta_2)=\int_{\cos\theta_2}^{\cos\theta_1}\big[C^{XY}(\theta)\big]^2 d(\cos\theta)\,,
\end{equation}
where $X,Y\in\{T,E\}$, and $C^{XY}(\theta)\equiv\langle X(\hat{\boldsymbol n}_1)Y(\hat{\boldsymbol n}_2)\rangle$ is the two-point correlation function averaged over all pixel pairs with $\hat{\boldsymbol n}_1\cdot\hat{\boldsymbol n}_2=\cos\theta$. The statistic $S_{1/2}^{XY}$ is typically evaluated over angular scales from $\theta_1=60^\circ$ to $\theta_2=180^\circ$, where the $p$-value is roughly minimized for the temperature data. For \textit{Planck} polarization data, no significant outliers are found, and the $p$-values show considerable variations between the component-separation methods employed \cite{Planck:2019evm}.

Given that a direct calculation of $C(\theta)$ is susceptible to noise bias, we derive it from the angular power spectrum via the following relation:
\begin{equation}
    C^{XY}(\theta)=\sum_{\ell=2}^{\ell_{\max}}\frac{2\ell+1}{4\pi}C_\ell^{XY}P_\ell(\cos\theta)\,,
\end{equation}
where $P_\ell$ are the Legendre polynomials. Thus, $S_{1/2}^{XY}$ can be written as \cite{Copi:2008hw}:
\begin{equation}
  S_{1/2}^{XY}=\sum_{\ell=2}^{\ell_{\max}}\sum_{\ell'=2}^{\ell_{\max}}\frac{2\ell+1}{4\pi}\frac{2\ell'+1}{4\pi}C_\ell^{XY}I_{\ell\ell'}C_{\ell'}^{XY}\,,
\end{equation}
where
\begin{equation}
    I_{\ell\ell'}\equiv\int_{-1}^{1/2}P_\ell(x)P_{\ell'}(x)dx\,.
\end{equation}
We compute the $S_{1/2}^{XY}$ statistic using \texttt{xQML} $EE$ and $TE$ power spectra with $\ell_{\max}=32$. We have verified that the results are insensitive to the choice of $\ell_{\max}$ in the range from 10 to 32. Although power spectra derived from masked $E$-mode fields would suffer from mode-mode coupling and $E$-$B$ mixing \cite{Zaldarriaga:1996xe,Yoho:2015bla}, we have observed that these effects are negligible for the $S_{1/2}^{EE}$ statistic under the configurations considered in this work (see Table~\ref{tab:noiseless-res}).
We also note that, for $E$-mode polarization, the range $(\theta_1, \theta_2)$ that minimizes the $p$-value is unknown, and we adopt the same range as used for temperature to avoid the selection effects.

\subsection{Quadrupole-octopole alignment}
\label{sec:qoa}
A significant alignment between the orientations of the quadrupole ($\ell=2$) and octopole ($\ell=3$) of CMB temperature anisotropies was discovered in the first-year WMAP data \cite{Tegmark:2003ve,deOliveira-Costa:2003utu,Schwarz:2004gk} at about 1.5\% significance level. Subsequent analyses based on \textit{Planck} data have consistently supported the presence of this anomaly \cite{Planck:2013lks,Copi:2013jna,Schwarz:2015cma}. Several approaches have been employed to characterize the multipole orientations: Ref.~\cite{deOliveira-Costa:2003utu} determines the direction $\hat{\boldsymbol n}$ that maximizes the angular momentum dispersion of a given $\ell$, $\sum_{m=-\ell}^{\ell} m^2|a_{\ell m}(\hat{\boldsymbol n})|^2$, thereby identifying the axis perpendicular to the multipole power's dominant plane. Here $a_{\ell m}(\hat{\boldsymbol n})$ denotes the spherical harmonic coefficients of the CMB map in a rotated coordinate system with its $z$-axis in the $\hat{\boldsymbol n}$-direction. Moreover, the octopole was found to be unusually planar, i.e., dominated by $m=\pm\ell$ coefficients, with a $p$-value of $\sim5\%$ \cite{deOliveira-Costa:2003utu}. 

An alternative method \cite{Schwarz:2004gk,Copi:2003kt} utilizes the Maxwell multipole vector decomposition, which represents the $\ell$-moment by $\ell$ unit vectors $\hat{\boldsymbol v}^{\ell;j}$ with $j=1,\dots,\ell$ and an overall amplitude. The quadrupole directions can be represented by two multipole vectors, and the octopole by three multipole vectors. The plane determined by any pair of multipole vectors at a given $\ell$ can be expressed as their oriented area \cite{Copi:2003kt},
\begin{equation}
   {\boldsymbol w}^{\ell;(i,j)} \equiv \hat{\boldsymbol v}^{\ell;i} \times \hat{\boldsymbol v}^{\ell;j}\,,
\end{equation}
whose magnitude is the area of the parallelogram spanned by the two vectors. The three oriented-area vectors of the octopole and the single oriented-area vector of the quadrupole are anomalously aligned, with $p$-value of about $0.2\%$ to $2\%$ \cite{Copi:2013jna}. The $S_{QO}$ statistic is defined to quantify the alignment of quadrupole and octopole area vectors:
\begin{equation}
    S_{QO} = \frac{1}{3}\sum_{i=1}^2\sum_{j=i+1}^3|{\boldsymbol w}^{2;(1,2)}\cdot{\boldsymbol w}^{3;(i,j)}|\,.
\end{equation}
We apply the estimator to the NILC $E$-mode maps after removing the monopole and dipole. In general, full-sky maps are preferred for recovering multipole vectors, as mode coupling induced by masking can degrade the reconstruction of low-$\ell$ multipoles \cite{Copi:2005ff,Planck:2013lks}. For AliCPT and its combination with SO, we use maps restricted to their respective scanning regions, although this introduces a $\sim0.8\sigma$ bias for AliCPT-alone maps, as shown in Table~\ref{tab:tot-res}.

\subsection{Point-parity asymmetry}
\label{sec:par}
The CMB temperature can be decomposed into even ($+$) and odd ($-$) parity components:
\begin{equation}
    T^{\pm}(\hat{\boldsymbol n})=\frac{1}{2}\left[T(\hat{\boldsymbol n})\pm T(-\hat{\boldsymbol n})\right]\,,
\end{equation}
which correspond to spherical harmonic modes with even and odd $\ell$, respectively. On the largest scales $2 \lesssim \ell \lesssim 30$, corresponding to the Sachs-Wolfe plateau of the temperature power spectrum, the even- and odd-parity contributions are expected to yield similar amplitudes. Contrarily, a preference for odd parity was identified in the \textit{WMAP} data \cite{Land:2005jq,Kim:2010gf,Gruppuso:2010nd} and later verified in the \textit{Planck} temperature observations \cite{Planck:2019evm} with a lower-tail probability of $\sim1\%$. In addition, it was found that this parity asymmetry is directional dependent, and the preferred axis strongly correlates with the preferred axes of the CMB kinematic dipole, quadrupole, and octopole \cite{P1,P2,P3}. Again, no anomalous probability was found in the \textit{Planck} polarization data \cite{Planck:2019evm}. The verification employed an estimator constructed from the even- and odd-$\ell$ power spectra:
\begin{equation}\label{eq:R_lmax}
    R^{XY}(\ell_{\max})=\frac{C^{XY}_{+}(\ell_{\max})}{C^{XY}_{-}(\ell_{\max})}\,,
\end{equation}
where
\begin{equation}
    C^{XY}_{+,-}(\ell_{\max})=\frac{1}{\ell_{\text{tot}}^{+,-}}\sum_{\ell=2,\ell_{\max}}^{+,-}\frac{\ell(\ell+1)}{2\pi}C_\ell^{XY}\,,
\end{equation}
with $\ell_{\text{tot}}^{+,-}$ being the number of even ($+$) or odd ($-$) multipoles in the range $[2, \ell_{\max}]$. 

Another estimator was used for \textit{Planck} polarization data:
\begin{equation}
    D^{XY}(\ell_{\max})=C^{XY}_{+}(\ell_{\max}) - C^{XY}_{-}(\ell_{\max})\,,
\end{equation}
since the low signal-to-noise ratio may lead to negative $C_{\ell}^{EE}$ or $C_{\ell}^{TE}$ estimates, particularly at low $\ell$, causing numerical problems in Eq.~\eqref{eq:R_lmax} when the ratio's denominator approaching zero (this occurs for $TE$ even in full-sky pure-CMB simulations due to cosmic variance). We thus calculate only $D^{XY}(\ell_{\max})$ from \texttt{xQML} $EE$ and $TE$ power spectra with $\ell_{\max}=24$, noting that varying $\ell_{\max}$ from 20 to 30 yields consistent results.

\section{Results}
\label{sec:rest}

In Section~\ref{sec:dm-val}, we validate the estimator for dipole modulation using modulated simulations with an input dipole in $E$ modes, and forecast the AliCPT's capability to identify this dipole from $E$-mode data. We then present the results of all four anomaly estimators on unconstrained isotropic simulations in Section~\ref{sec:tst-unc}.

\subsection{Dipole modulation: validation on modulated simulations}
\label{sec:dm-val}

\begin{figure}[tbp]
    \centering
    \begin{subfigure}[b]{0.48\textwidth}
        \centering
        \includegraphics[width=\textwidth]{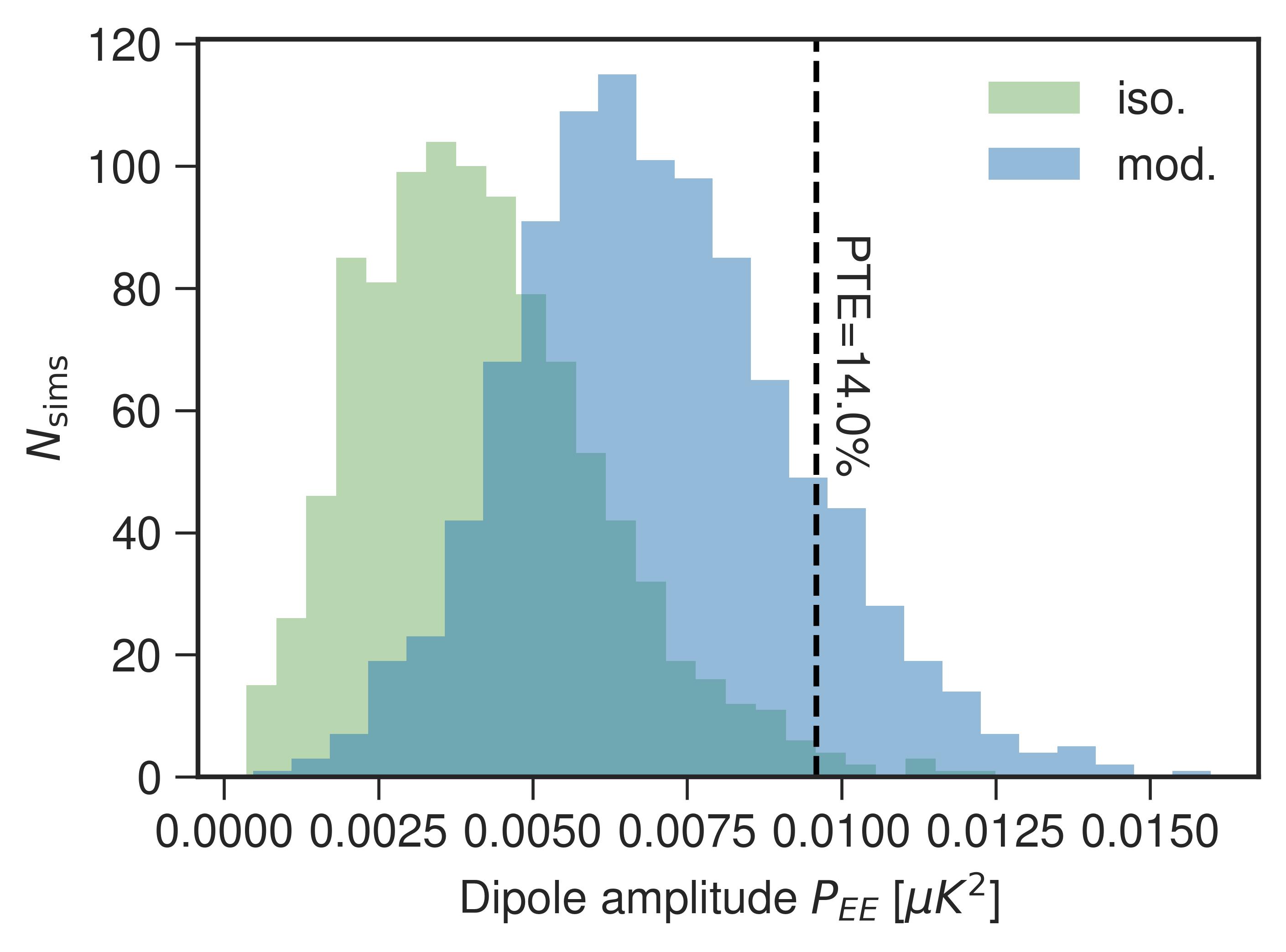}
    \end{subfigure}
    \begin{subfigure}[b]{0.48\textwidth}
        \centering
        \includegraphics[width=\textwidth]{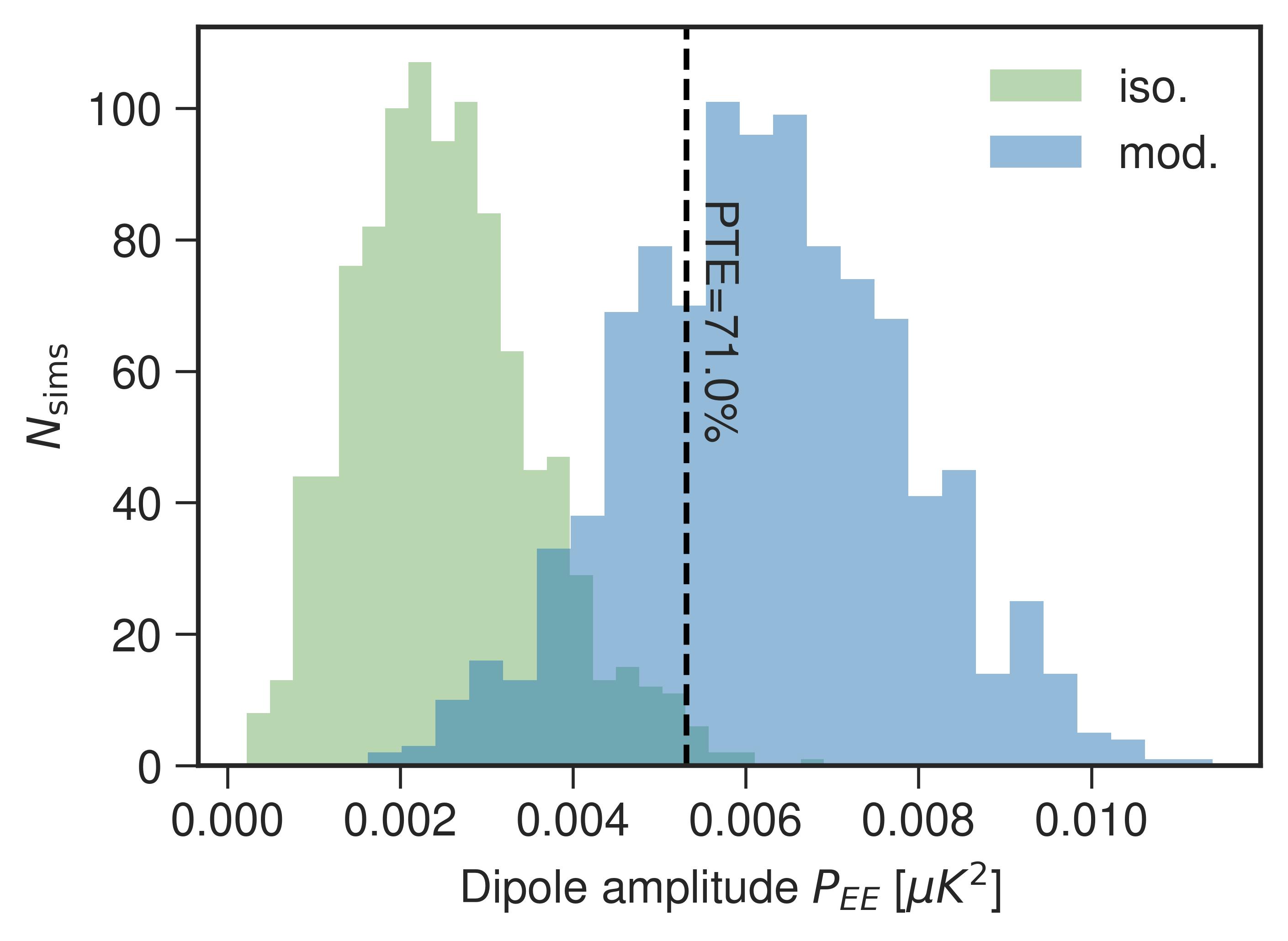}
    \end{subfigure}
    \begin{subfigure}[b]{0.48\textwidth}
        \centering
        \includegraphics[width=\textwidth]{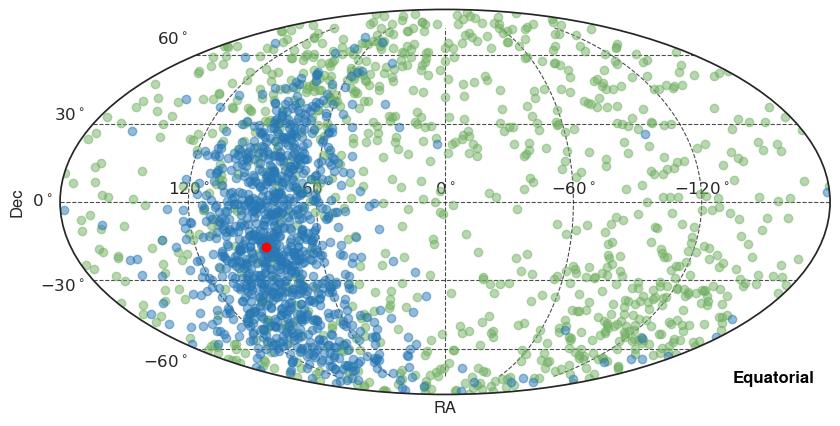}
    \end{subfigure}
    \begin{subfigure}[b]{0.48\textwidth}
        \centering
        \includegraphics[width=\textwidth]{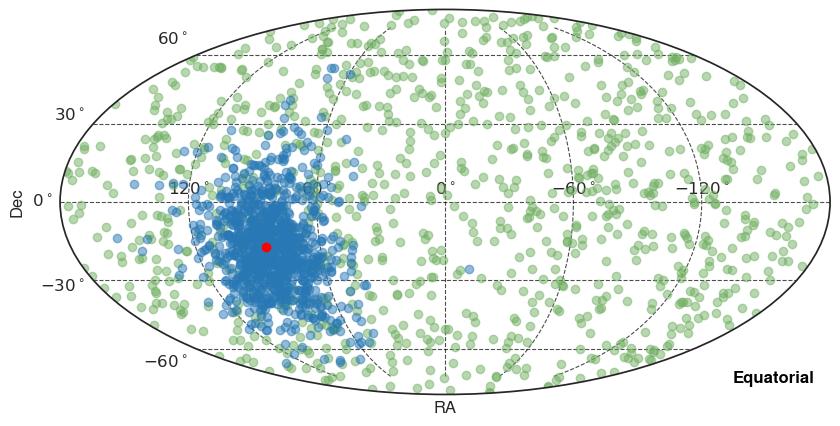}
    \end{subfigure}
    \begin{subfigure}[b]{0.48\textwidth}
        \centering
        \includegraphics[width=\textwidth]{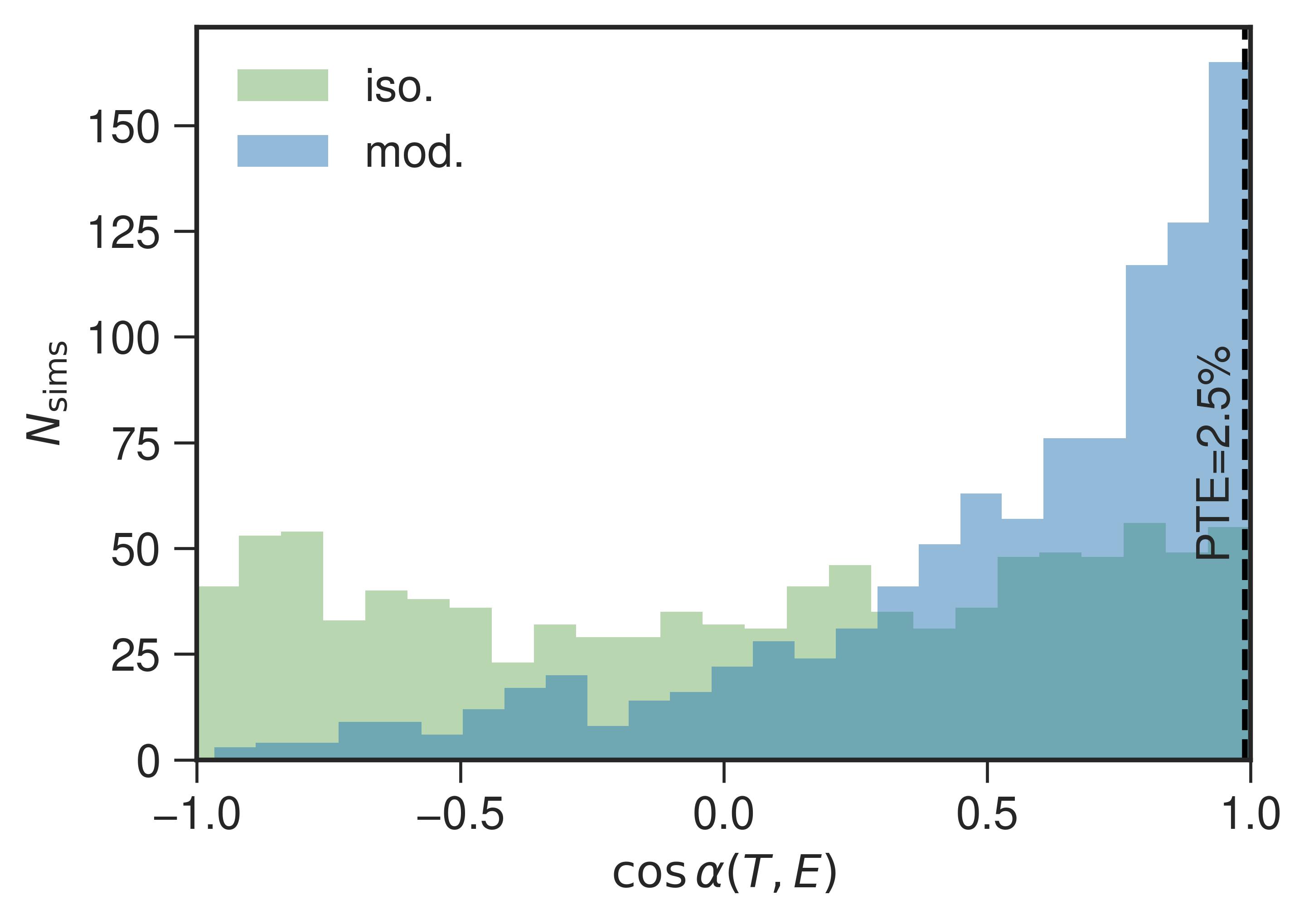}
    \end{subfigure}
    \begin{subfigure}[b]{0.48\textwidth}
        \centering
        \includegraphics[width=\textwidth]{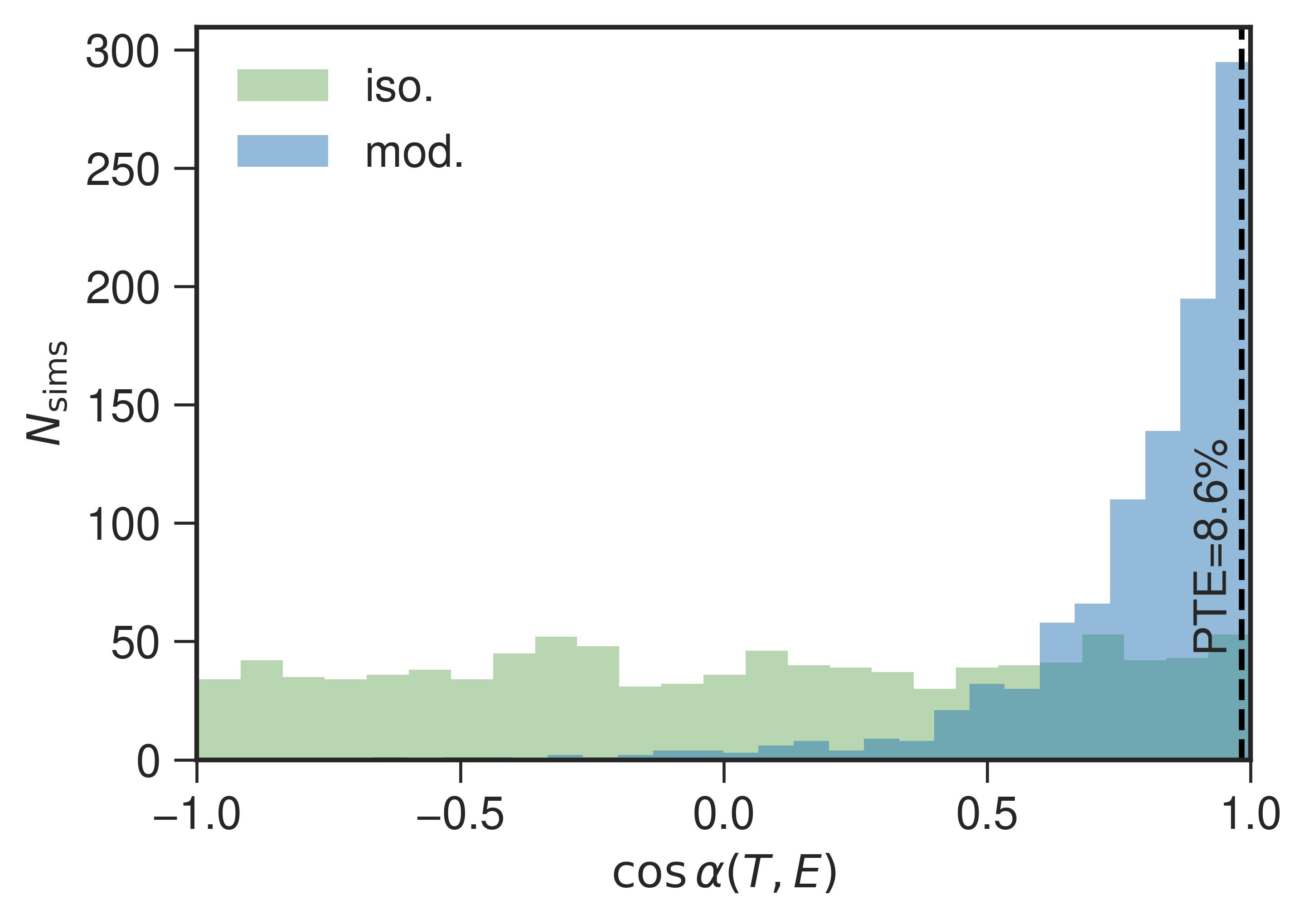}
    \end{subfigure}

    \caption{\textit{Top}: Distributions of the $E$-mode dipole amplitude derived from LVMs of the dipole-modulated (blue) and isotropic (green) simulations, for goal-AliCPT (left) and AliCPT+SO (right) cases. The disc radius is set to $r=4^\circ$. \textit{Middle}: Distributions of the measured $E$-mode dipole direction. The input dipole direction, $(\mathrm{RA}, \mathrm{DEC}) = (86^\circ, -17^\circ)$, is labeled as a red dot. \textit{Bottom}: Distributions of the cosine of the angle between the temperature and $E$-mode dipole directions.}
    \label{fig:distr-dm}
\end{figure}

\begin{figure}[tbp]
    \centering
    \begin{subfigure}[b]{0.7\textwidth}
        \centering
        \includegraphics[width=\textwidth]{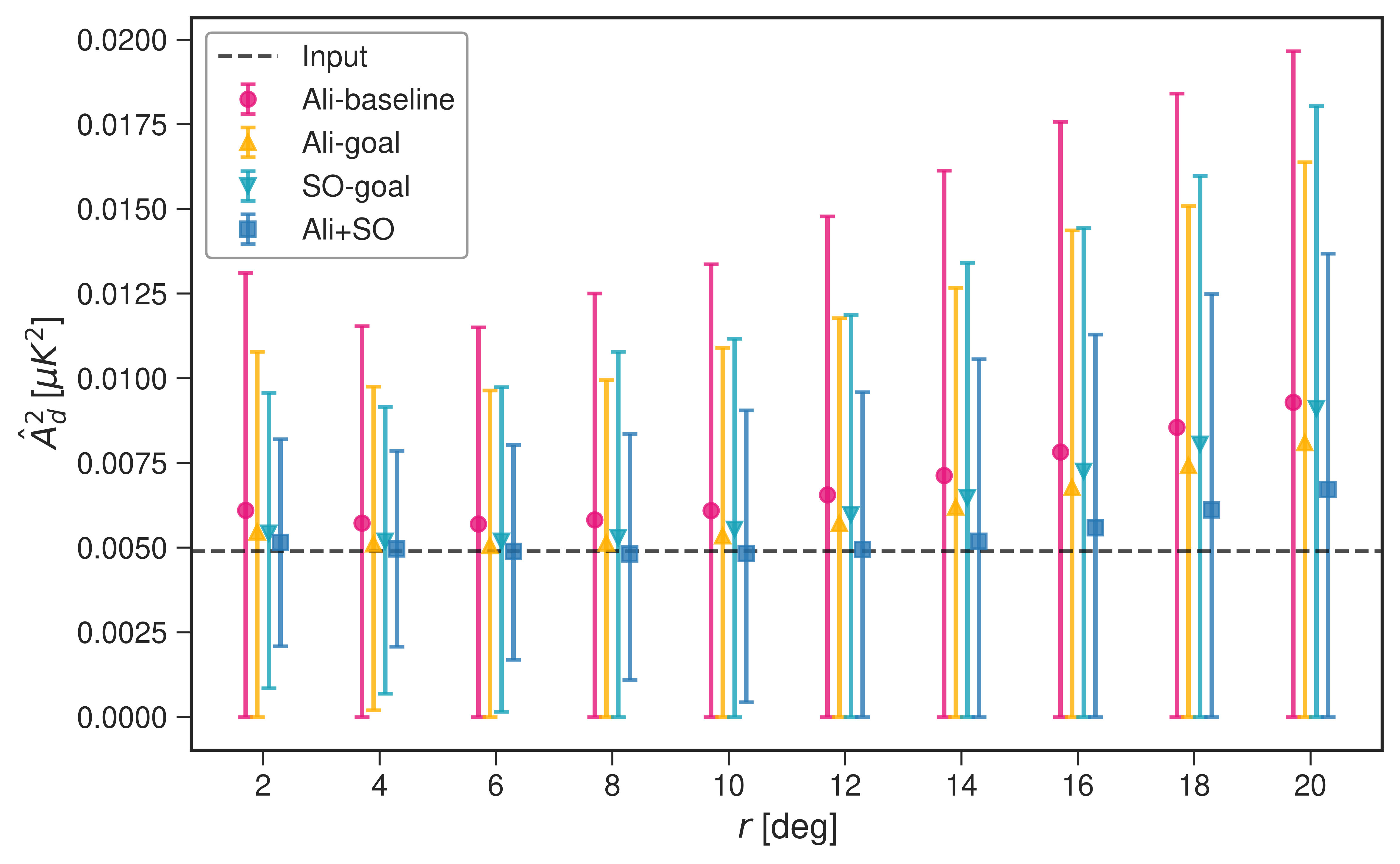}
    \end{subfigure}
    \caption{The square of the $E$-mode dipole amplitude estimated from dipole-modulated simulations, as a function of the disc radius $r$, for baseline-AliCPT (pink), goal-AliCPT (yellow), \revise{goal-SO (cyan),} and AliCPT+SO (blue) sensitivities. The dashed line shows the input dipole amplitude, $A_d^2=0.049$.}
    \label{fig:dm-amp2}
\end{figure}

\begin{table}[tbp]
    \centering
    \caption{Mean and 1$\sigma$ interval of the $E$-mode dipole modulation statistics with $r=4^\circ$ estimated from modulated simulations. The values in parentheses in the first line show the 1$\sigma$ uncertainties for the isotropic simulations. The ``ideal'' indicates the full-sky and noiseless case. The two PTEs represent the fraction of modulated simulations for which $P_{EE}$ or $\cos\alpha_{TE}$ exceeds the 99th percentile of the isotropic simulations.}
    \renewcommand\arraystretch{1.1}
    \small  
    \setlength{\tabcolsep}{1pt}  
    \begin{tabular}{l |ccccc}
        \hline
        \hline
        Estimator&Ali-baseline&Ali-goal&SO-goal&Ali+SO&Ideal \\
          \hline
        $P_{EE}$ [$\times10^{-3}\mu{\rm K}^2$] & $7.8\pm2.9$ (2.5) & $7.0\pm2.3$ (2.0) & $7.0\pm2.1$ (1.8) & $6.2\pm1.6$ (1.0) & $6.1\pm1.2$ (0.8)\\
        PTE (99\% iso.) & 4.1\% & 14.0\% & 15.4\% &71.0\% & 96.9\%\\
        $\hat A_d^2$ [$\times10^{-3}$] & $5.4\pm5.6$ & $5.1\pm4.8$ & $5.2\pm4.2$ & $4.9\pm2.9$ & $4.9\pm2.1$\\
        (RA, DEC)& $(85^\circ,-16^\circ)\pm22^\circ$ & $(86^\circ,-17^\circ)\pm19^\circ$ & $(86^\circ,-17^\circ)\pm16^\circ$ & $(86^\circ,-18^\circ)\pm12^\circ$ & $(86^\circ,-17^\circ)\pm8^\circ$  \\
        $\cos\alpha_{TE}$ & $0.63^{+0.29}_{-0.64}$ & $0.68^{+0.25}_{-0.56}$ & $0.76^{+0.18}_{-0.40}$ & $0.86^{+0.10}_{-0.23}$ & $0.92^{+0.06}_{-0.16}$\\
        PTE (99\% iso.) & 3.2\% & 2.5\% & 3.0\% & 8.6\% & 19.5\%\\
      \hline 
    \end{tabular}
    \label{tab:dm-res}
\end{table}

We first validate the local variance estimator using 1000 CMB realizations modulated by a dipole with amplitude $A_d = 0.07$ and direction $(l, b) = (221^\circ, -22^\circ)$ in Galactic coordinates---the best-fit values from the \textit{Planck} 2018 temperature data measured with the QML estimator \cite{Planck:2019evm}. This direction is converted to Equatorial coordinates $(\mathrm{RA}, \mathrm{DEC}) = (86^\circ, -17^\circ)$ to match our maps and masks. The modulation is implemented in harmonic space (see Appendix~\ref{app:dm-harm}) and is equivalent to applying Eq.~\eqref{eq:dm} in real space. Following the Planck team's configuration \cite{Planck:2019evm}, we compute the LVMs at $N_{\text{side}}=16$ resolution (pixel size $\theta_{\rm pix}=3.66^\circ$) for both modulated and isotropic realizations at $N_{\text{side}}=64$, with the circular disc radius $r$ ranging from $2^\circ$ to $20^\circ$. \textit{Planck} selected $r=4^\circ$ as it ensures that every pixel on the sky map is covered by at least one disc, while the local variances are slightly correlated between adjacent discs \cite{Sanyal:2024iyv}. We then derive the best-fit dipole from the un-normalized and normalized LVMs using Eqs.~\eqref{eq:dm-unnorm} and \eqref{eq:dm-norm}, respectively. As expected, we find that the dipole directions obtained from the two are consistent.

The resulting estimates of the $E$-mode dipole modulation amplitude and direction for modulated simulations with $r=4^\circ$ are summarized in Table~\ref{tab:dm-res}, and their distributions are plotted in Fig.~\ref{fig:distr-dm}. As a reference, the $E$-mode dipole amplitude from \textit{Planck} SMICA data is $P_{EE}=0.008\pm0.004\,\mu\mathrm{K}^2$ \cite{Shi:2022hxc}. From Table~\ref{tab:dm-res}, we observe that as the sensitivity improves from the baseline to the goal and further to the AliCPT+SO combination, the bias and error of the un-normalized amplitude $P_{EE}$ decrease, approaching the ideal full-sky noiseless case. The probability to exceed (PTE) the $99\%$ upper limit of isotropic simulations increases from $4.1\%$ (AliCPT-baseline) to $14.0\%$ (AliCPT-goal) and $71.0\%$ (AliCPT+SO), indicating that the combined AliCPT+SO dataset can robustly detect the injected dipole modulation with a probability of $71.0\%$. \revise{The standalone SO‑goal configuration achieves a PTE of $15.4\%$, similar to that of AliCPT‑goal despite SO's larger sky coverage. This can be understood by noting that the direction of the injected dipole is located in the southern hemisphere. The AliCPT footprint covers the northern sky where the dipole field has the largest gradient \cite{Shi:2022hxc}, making AliCPT particularly sensitive to the dipole modulation.}

All recovered dipole directions are consistent with the input, and the $1\sigma$ angular scatter shrinks from $22^\circ$ (AliCPT-baseline) to $12^\circ$ (AliCPT+SO) (see Table~\ref{tab:dm-res}). The cosine of the angle between the dipole directions recovered from temperature and $E$-mode maps, $\cos\alpha_{TE}$, also increases with a PTE from $3.2\%$ (AliCPT-baseline) to $8.6\%$ (AliCPT+SO). Interestingly, the inferred $\cos\alpha_{TE}$ from the \textit{Planck} $T$- and $E$-mode data ranges from 0.86 to 0.99, with $p$-values varying between 6.9\% to 0.9\% depending on the component separation methods employed \cite{Planck:2019evm}. Even for the ideal case, assuming a dipole amplitude of 0.07 for both temperature and polarization, such a high alignment is unusual (99\% PTE $\sim19.5\%$).

Figure~\ref{fig:dm-amp2} displays the squared dipole amplitude $\hat{A}_d^2$ estimated from modulated simulations as a function of disc radius $r$, for different noise configurations. The dashed line indicates the input amplitude $A_d^2=0.049$. For small $r$ ($\lesssim 4^\circ$), the uncertainties increase due to insufficient pixels per disc to reliably compute the variance; for large $r$ ($\gtrsim 10^\circ$), the estimates become biased because the localization of LVMs breaks down, and thus the variance can no longer be separated from the dipole background term in Eq.~\eqref{eq:lvm}. The optimal radius is around $r=4^\circ$ to $8^\circ$, where the recovered amplitude for goal-AliCPT and AliCPT+SO schemes is unbiased.

In summary, the local variance estimator successfully recovers the injected dipole modulation in $E$ modes. With the goal sensitivity, AliCPT alone can detect the 7\% modulation at a modest significance (PTE $\sim14\%$), and the combination with SO LAT, nearly doubling the sky coverage, significantly boosts the detection power (PTE $\sim71\%$). This demonstrates that AliCPT+SO will provide a powerful test of the dipolar power asymmetry in CMB polarization. \revise{The detection power of AliCPT+SO is not driven solely by SO, given that the PTE of SO-goal ($\sim15\%$) is comparable to that of AliCPT-goal.}

\subsection{Tests on unconstrained isotropic simulations}
\label{sec:tst-unc}

\begin{figure}[tbp]
    \centering
    \begin{subfigure}[b]{0.45\textwidth}
        \centering
        \includegraphics[width=\textwidth]{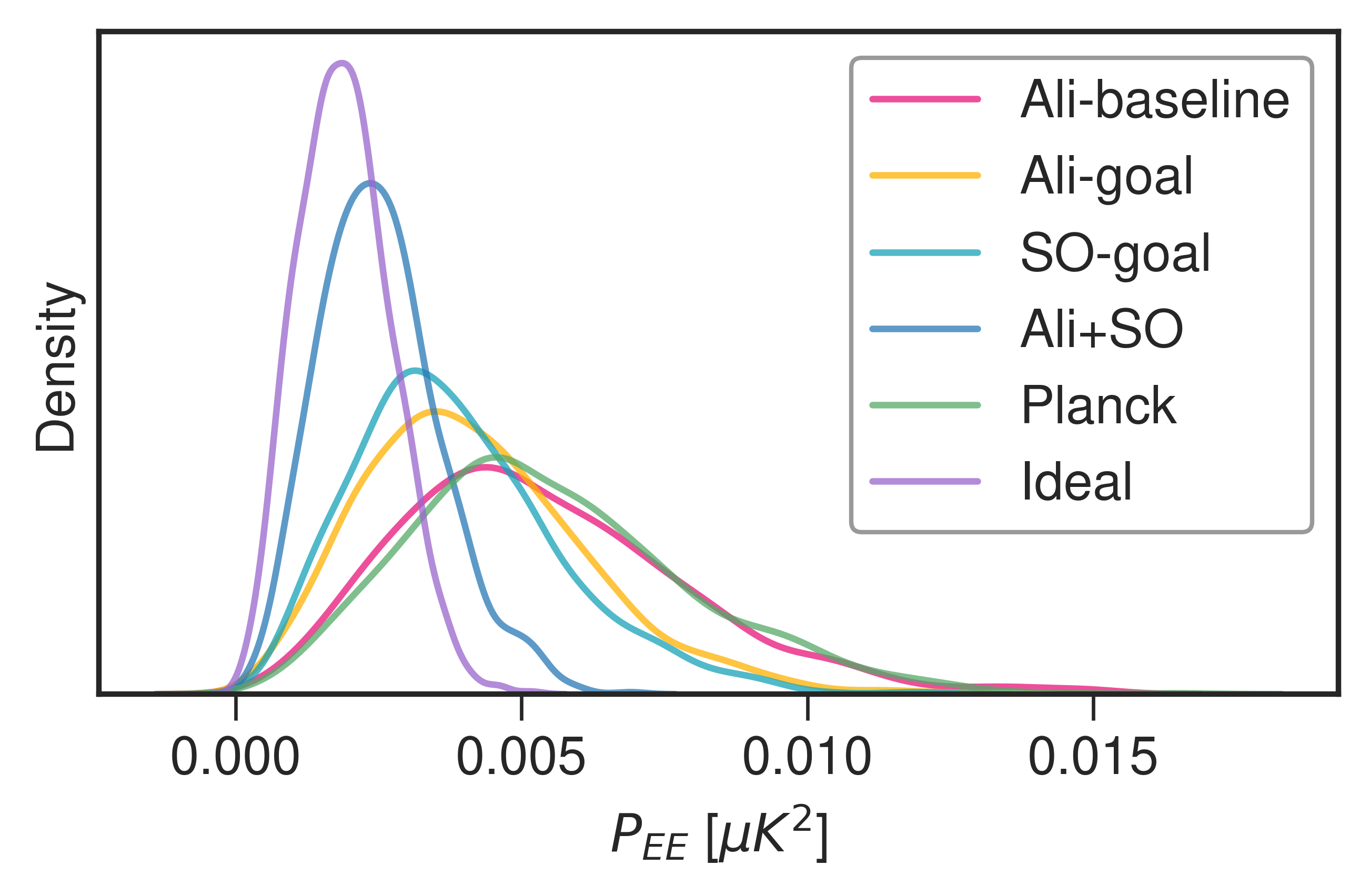}
    \end{subfigure}
    \begin{subfigure}[b]{0.45\textwidth}
        \centering
        \includegraphics[width=\textwidth]{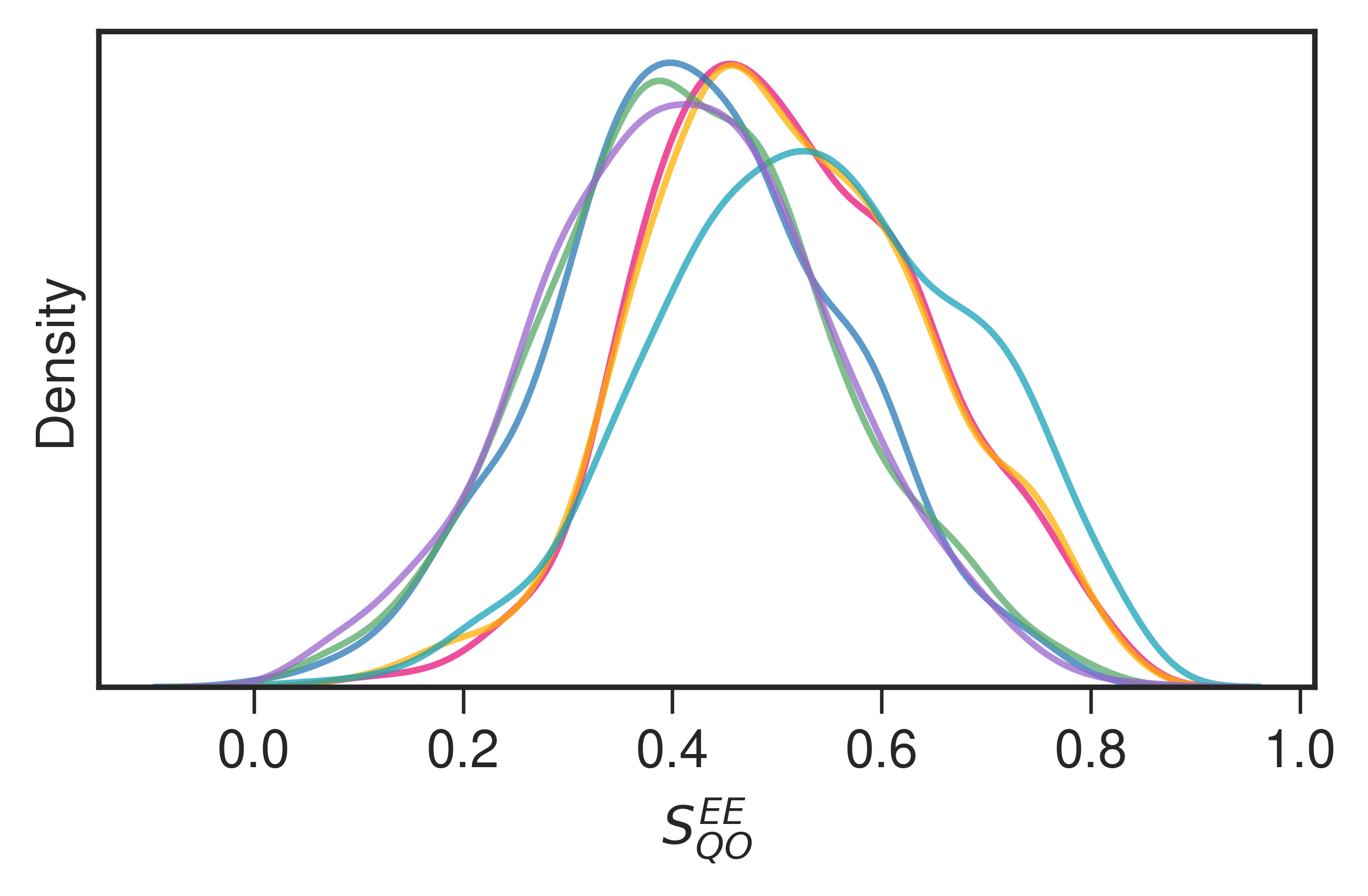}
    \end{subfigure}
    \begin{subfigure}[b]{0.45\textwidth}
        \centering
        \includegraphics[width=\textwidth]{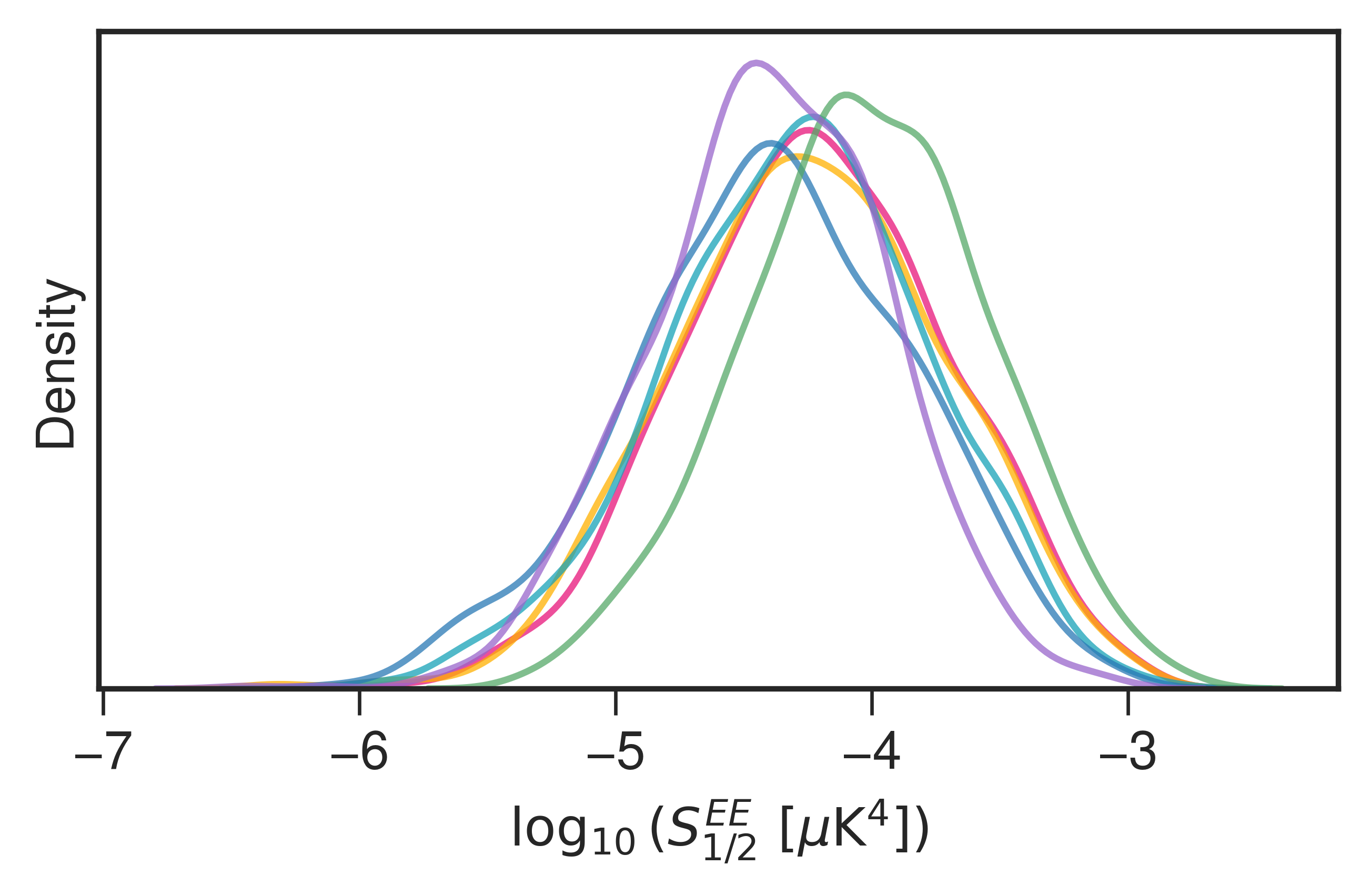}
    \end{subfigure}
    \begin{subfigure}[b]{0.45\textwidth}
        \centering
        \includegraphics[width=\textwidth]{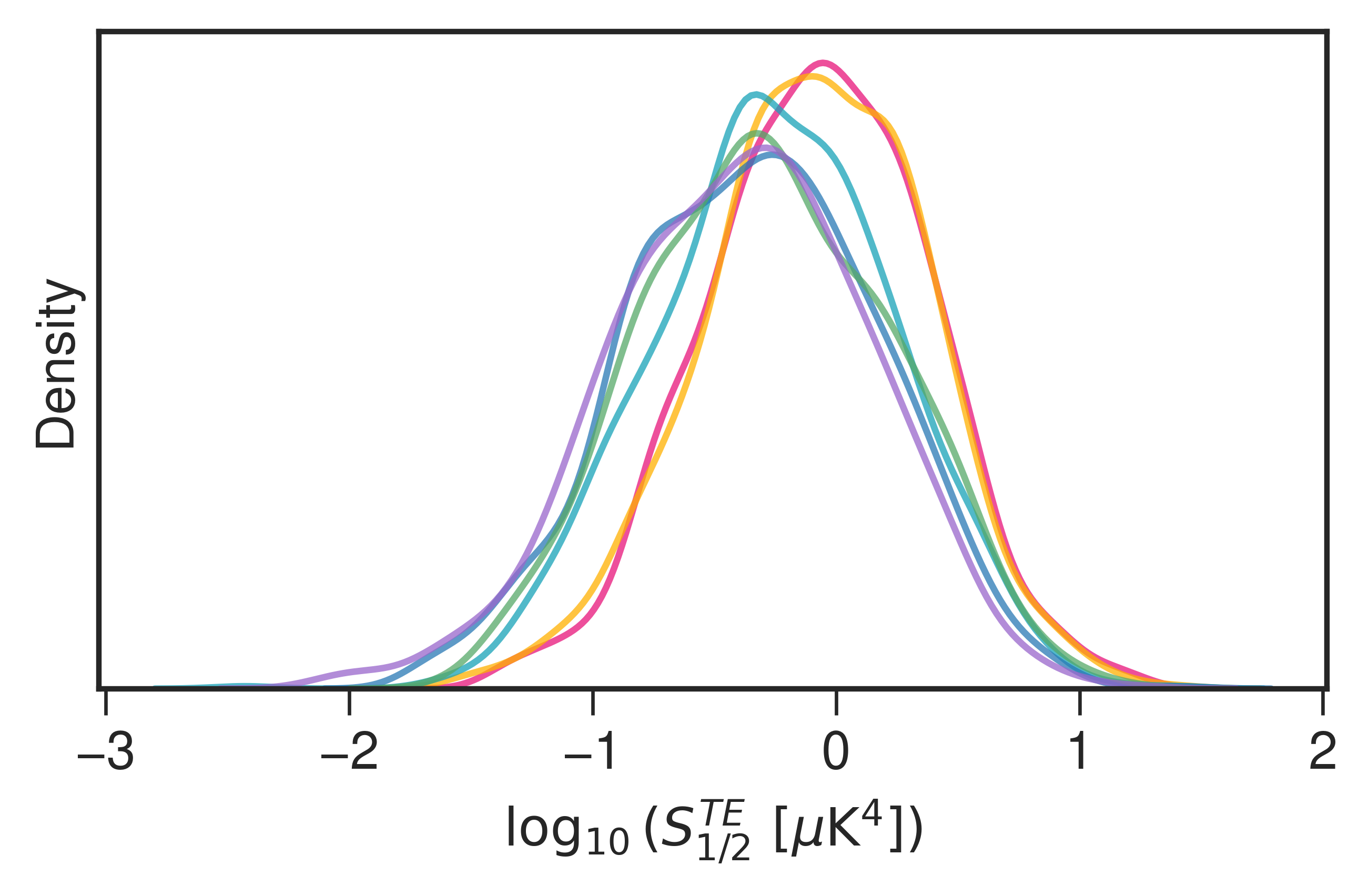}
    \end{subfigure}
    \begin{subfigure}[b]{0.45\textwidth}
        \centering
        \includegraphics[width=\textwidth]{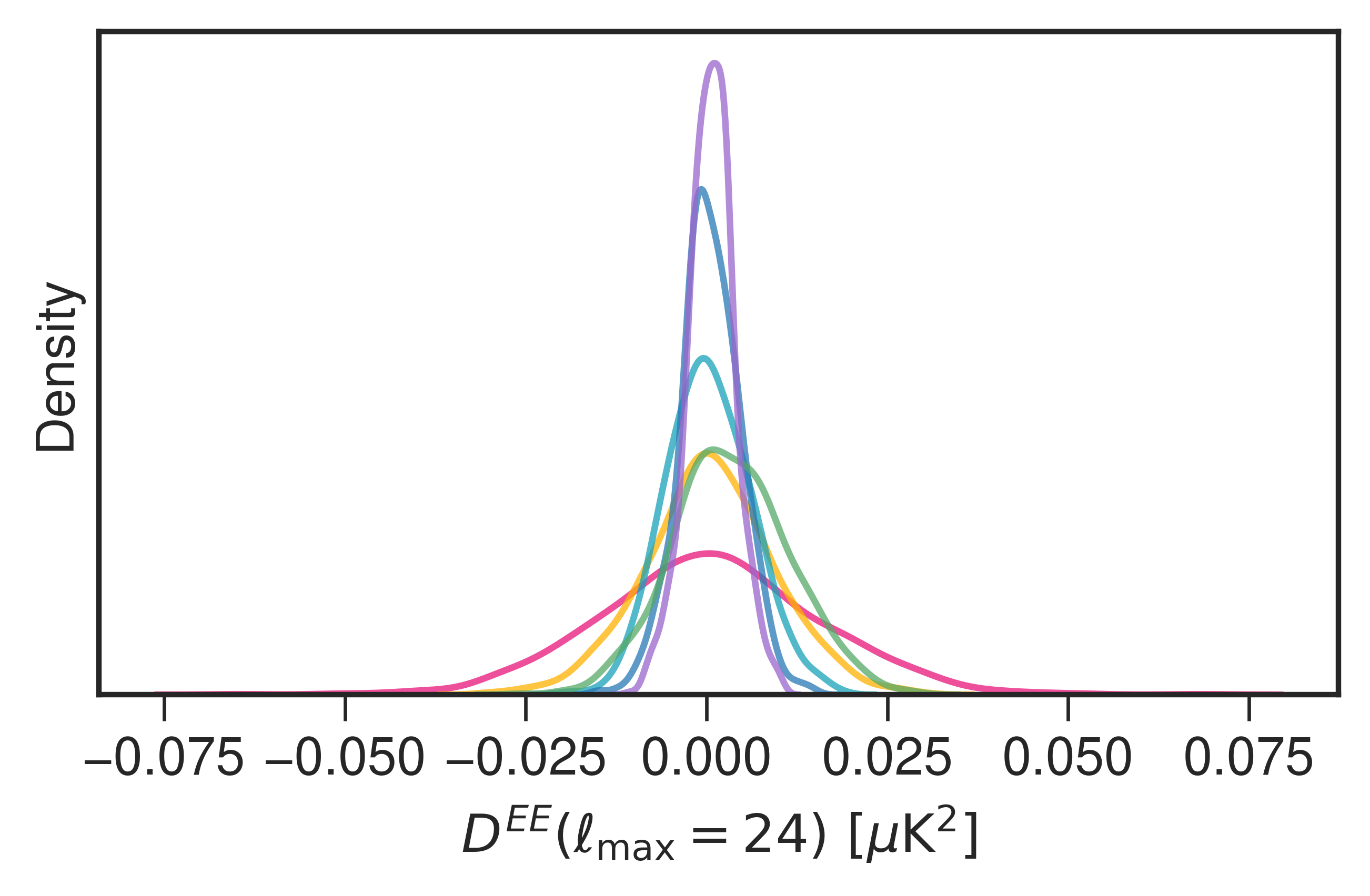}
    \end{subfigure}
    \begin{subfigure}[b]{0.45\textwidth}
        \centering
        \includegraphics[width=\textwidth]{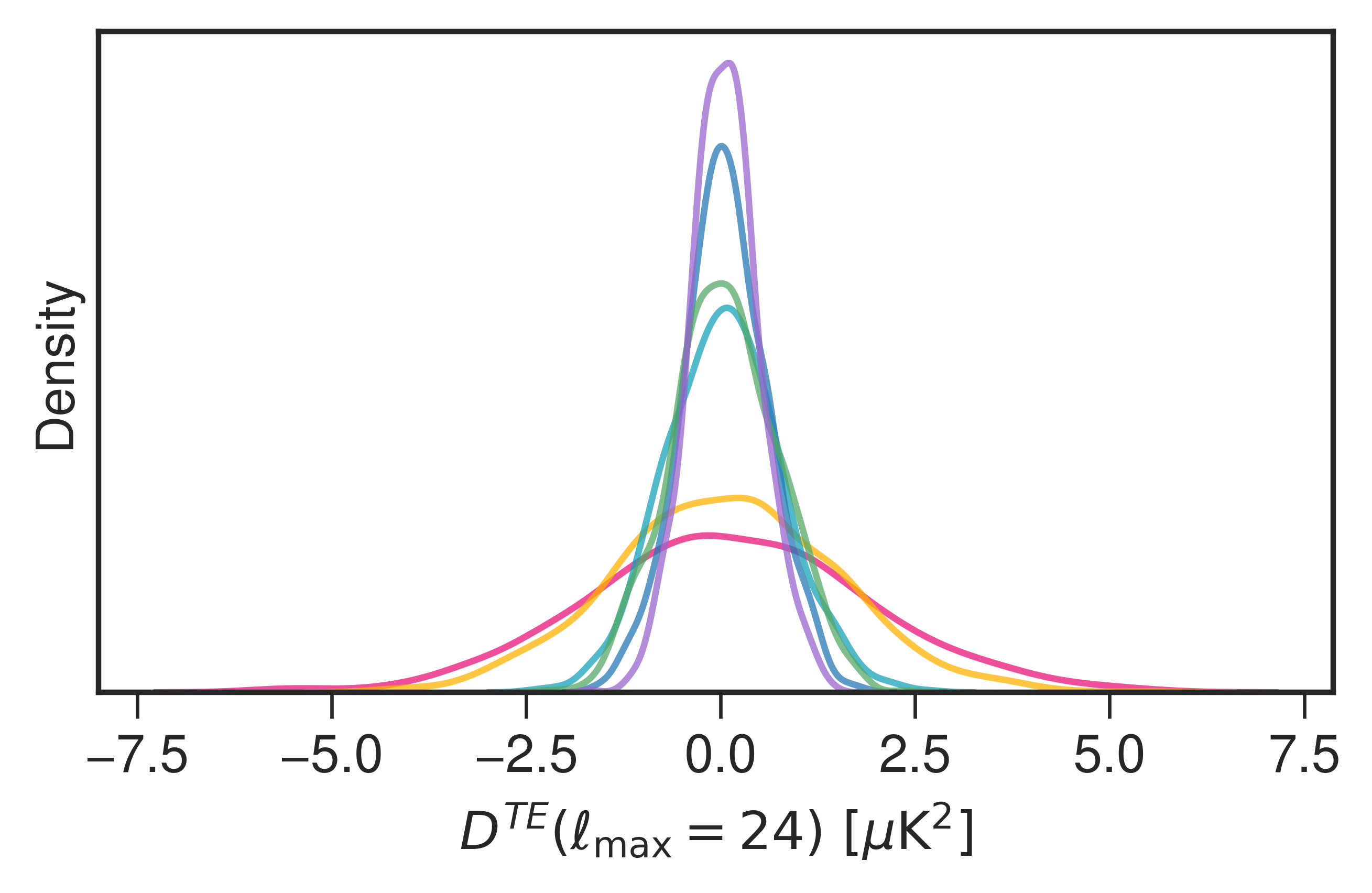}
    \end{subfigure}

    \caption{Density distributions of the $E$-mode anomaly statistics derived from unconstrained simulations, for baseline-AliCPT (pink), goal-AliCPT (yellow), \revise{goal-SO (cyan),} AliCPT+SO (blue), \textit{Planck} SMICA noise (green), and an ideal case (purple) representing scattering owing to cosmic variance.}
    \label{fig:distr-all}
\end{figure}

\begin{table}[tbp]
    \centering
    \caption{Mean and $1\sigma$ interval of the $E$-mode anomaly estimators derived from unconstrained simulations. The ``ideal'' in the last column indicates the full-sky and noiseless case.}
    \renewcommand\arraystretch{1.3}
    \small  
    \setlength{\tabcolsep}{4pt}  
    \begin{tabular}{l |cccccc}
        \hline
        \hline
        Estimator&Ali-baseline&Ali-goal&SO-goal&Ali+SO&Planck&Ideal \\
          \hline
        $P_{EE}$ [$\times10^{-3}\mu{\rm K}^2$]& $5.4\pm2.5$ & $4.2\pm2.0$ & $3.9\pm1.8$ & $2.5\pm1.0$ & $5.5\pm2.4$ & $1.9\pm0.8$ \\
        \hline
        $\log_{10}(S_{1/2}^{EE}$ $[\mu{\rm K}^4])$ & $-4.22^{+0.54}_{-0.51}$ & $-4.24^{+0.55}_{-0.53}$ & $-4.28^{+0.49}_{-0.52}$ & $-4.39^{+0.56}_{-0.53}$ & $-4.03^{+0.50}_{-0.48}$ & $-4.40^{+0.44}_{-0.49}$ \\
        $\log_{10}(S_{1/2}^{TE}$ $[\mu{\rm K}^4])$ & $-0.07^{+0.44}_{-0.46}$ & $-0.07^{+0.43}_{-0.47}$ & $-0.25^{+0.49}_{-0.50}$ & $-0.33^{+0.53}_{-0.52}$ & $-0.30^{+0.55}_{-0.51}$ & $-0.39^{+0.53}_{-0.53}$ \\
        \hline
        $D^{EE}(24)$ [$\times10^{-3}\mu{\rm K}^2$] & $1\pm14$ & $0\pm9$ & $0\pm6$ & $0\pm4$ & $3\pm8$ & $1\pm3$ \\
        $D^{TE}(24)$ [$\mu{\rm K}^2$] & $0.09\pm1.75$ & $0.04\pm1.39$ & $0.04\pm0.72$ & $0.04\pm0.53$ & $0.04\pm0.69$ & $0.05\pm0.43$ \\
        \hline
        $S_{QO}^{EE}$ & $0.51\pm0.13$ & $0.51\pm0.14$ & $0.53\pm0.15$ & $0.42\pm0.14$ & $0.42\pm0.14$ & $0.41\pm0.14$ \\

      \hline 
    \end{tabular}
    \label{tab:tot-res}
\end{table}

\begin{table}[tbp]
    \centering
    \caption{Mean and $1\sigma$ interval of the large-angle correlation statistics derived from nearly-noiseless simulations on different sky areas.}
    \renewcommand\arraystretch{1.3}
    \begin{tabular}{l |cccc}
        \hline
        \hline
        Estimator & AliCPT & SO & Ali+SO &Planck \\
          \hline
        $\log_{10}(S_{1/2}^{EE}$ $[\mu{\rm K}^4])$ & $-4.31^{+0.43}_{-0.48}$ & $-4.37^{+0.43}_{-0.47}$ & $-4.39^{+0.44}_{-0.49}$ & $-4.39^{+0.44}_{-0.49}$ \\
        $\log_{10}(S_{1/2}^{TE}$ $[\mu{\rm K}^4])$ & $-0.06^{+0.43}_{-0.44}$ & $-0.27^{+0.49}_{-0.47}$ & $-0.38^{+0.52}_{-0.53}$ & $-0.37^{+0.52}_{-0.52}$ \\
      \hline 
    \end{tabular}
    \label{tab:noiseless-res}
\end{table}

In this section, we apply the four anomaly estimators defined in Sections~\ref{sec:dm}--\ref{sec:par} to unconstrained simulations, i.e., standard $\Lambda$CDM realizations without any injected anomalies. Table~\ref{tab:tot-res} reports the mean and $1\sigma$ intervals of each statistic computed from 1000 unconstrained realizations for different experimental configurations: baseline-AliCPT, goal-AliCPT, \revise{goal-SO}, AliCPT+SO (both goal), \textit{Planck} SMICA (with the \textit{Planck} 2018 common mask for all estimators except $S_{QO}$, for which the full-sky maps are used), and an ideal full-sky noiseless case. Their distributions are shown in Fig.~\ref{fig:distr-all}. The following points summarize the key findings.

\begin{itemize}
  \item \textbf{Dipole modulation} ($P_{EE}$): The AliCPT baseline and \textit{Planck} show similar mean amplitudes and errors, both significantly above the ideal cosmic-variance-limited level, indicating that the current $E$-mode constraints are noise-dominated. The target sensitivity of AliCPT can reduce the error bar by about 25\%, \revise{similar to the SO-goal case}. As the sky coverage increases from AliCPT alone to the AliCPT+SO combination, the mean and error drop to near the ideal values, demonstrating the importance of large sky coverage for the dipole modulation measurements.
  \item \textbf{Lack of large-angle correlations} ($\log_{10}S_{1/2}$): The $EE$ and $TE$ correlation statistics exhibit different behaviors in distributions. We note that, although the uncertainties in $\log_{10}S_{1/2}^{EE}$ for AliCPT, SO, \textit{Planck}, AliCPT+SO, and the ideal case appear comparable, for this logarithm quantity with a fixed error bar, a smaller central value corresponds to a narrower absolute uncertainty in $S_{1/2}^{EE}$. Consequently, AliCPT+SO actually has the lowest uncertainty in $S_{1/2}^{EE}$ and $S_{1/2}^{TE}$ among the first five cases. We observe that, for $\log_{10}S_{1/2}^{EE}$, \textit{Planck} yields a positive deviation of $\sim1\sigma$ from the ideal value. This arises because the two‑point correlation function $C^{EE}(\theta)$ derived from noise‑dominated \textit{Planck} simulations oscillates with $\theta$, causing the integral of its square (i.e., $S_{1/2}^{EE}$) to shift to larger values and simultaneously raising the lower bound of the $1\sigma$ interval. 
  
  For $\log_{10}S_{1/2}^{TE}$, we find that its mean value varies significantly with sky coverage. To decouple the effects of noise and sky cuts on this statistic, we compute the statistical distributions on different sky areas from almost-noiseless simulations, i.e., with small white noise (2 $\mu$K-arcmin for $T$ and 0.02 $\mu$K-arcmin for $E$ modes) added; the results are shown in Table~\ref{tab:noiseless-res}. The mean and uncertainty of the $TE$ statistic are indeed more sensitive to sky areas than those of the $EE$ statistic. For small sky areas, the measured $TE$ power spectrum, even when limited by cosmic variance, fluctuates around zero at low multipoles, leading to an overall increase in $S_{1/2}^{TE}$. We therefore caution against comparing $S_{1/2}^{TE}$ across different sky areas.
  \item \textbf{Point-parity asymmetry} ($D(\ell_{\max}=24)$): 
  The $D^{EE}$ and $D^{TE}$ estimators are unbiased for all experimental configurations. However, due to limited sky coverage, their scatter is larger for AliCPT (baseline and goal cases) than for SO and \textit{Planck} SMICA simulations. \revise{For $D^{TE}$, the SO-goal configuration yields substantially smaller error bars than AliCPT-goal, benefiting from its lower noise level and larger sky fraction. The AliCPT+SO dataset reduces the scatter even further, approaching the ideal level.}
  \item \textbf{Quadrupole-octopole alignment} ($S_{QO}$): The $S_{QO}^{EE}$ statistic, which characterizes the alignment of $\ell=2$ and $\ell=3$ modes, is dominated by cosmic variance; thus, its uncertainty does not benefit from higher sensitivity (beyond that of \textit{Planck}). When the sky coverage is restricted (as in the AliCPT-alone cases), mode coupling introduces a positive bias of approximately $\sim0.8\sigma$. This bias is eliminated when the sky fraction reaches $\sim80\%$ (AliCPT+SO case). Hence, a large sky fraction is essential for this anomaly to manifest, whereas low noise is not a critical requirement.
\end{itemize}

Overall, the unconstrained simulations show that AliCPT alone suffers from modest biases and inflated errors due to its partial sky coverage. \revise{The SO LAT configuration exhibits higher sensitivity than AliCPT for $C_\ell^{TE}$-based estimators ($\log_{10}S_{1/2}^{TE}$ and $D^{TE}(\ell_{\max}=24)$), but performs comparably to AliCPT for other estimators ($P_{EE}$, $S_{QO}$, $\log_{10}S_{1/2}^{EE}$, and $D^{EE}(\ell_{\max}=24)$).} The combination of AliCPT and SO effectively restores the distributions to those under the ideal full-sky case, serving as a near-cosmic-variance-limited benchmark for future $E$-mode anomaly searches.

\section{Conclusions}
\label{sec:concl}

In this paper, we present $E$-mode polarization forecasts for the capability of the AliCPT experiment alone and in combination with the Simons Observatory (SO) LAT to test the large-scale anomalies persistently observed in CMB temperature maps. The complementary sky coverage of AliCPT and SO provides the large sky fraction essential for studying large-scale anomalies. Based on 1000 unconstrained simulations that include Galactic foregrounds, instrumental noise (baseline and goal sensitivities for AliCPT, goal for SO), and applying the NILC component separation method, we evaluate four well-known anomaly statistics: dipole modulation (hemispherical power asymmetry), lack of large-angle correlations, quadrupole-octopole alignment, and point-parity asymmetry.

In Section~\ref{sec:dm-val}, we validate the local variance estimator, a method of detecting dipole modulation, using modulated $E$-mode simulations with an input dipole amplitude $A_d = 0.07$ and direction matching the \textit{Planck} temperature best fit. For the AliCPT four years' sensitivity, the probability to exceed (PTE) the 99th percentile of isotropic simulations is 14\%, \revise{comparable to the SO goal case}, and rises to 71\% when AliCPT is combined with SO. This demonstrates that the combined AliCPT+SO dataset could reject the fluke hypothesis at the 99\% confidence level with a probability of 71\%, under the assumption of $A_d = 0.07$. Validation tests confirm that the local variance estimator produces an unbiased measurement of the modulation amplitude $A_d$ for disc radii $r \in [4^\circ, 8^\circ]$ under the goal sensitivity. The reconstructed dipole orientations remain consistent with the input direction in all cases, while the cosine of the angle between the $T$- and $E$-mode dipole directions, $\cos\alpha_{TE}$, shows only modest statistical significance (PTE = 19.5\%) even under the ideal full-sky scenario.

For the four anomalies considered, tests on unconstrained $\Lambda$CDM realizations indicate that AliCPT data alone may exhibit systematic offsets or increased uncertainties due to incomplete sky coverage, particularly for point‑parity asymmetry and quadrupole‑octopole alignment. Incorporating SO LAT observations largely overcomes this limitation, restoring the anomaly statistic distributions to approximate the ideal full‑sky level. \revise{The standalone SO-goal results reveal that SO contributes significantly to the two estimators based on $TE$ power spectra, whereas for other estimators the two experiments are almost equally important. Especially for dipole modulation, AliCPT, as the only northern-hemisphere ground-based CMB polarization experiment, captures the steep gradient of the dipole modulation field and effectively complements southern surveys by increasing the total usable sky fraction.} Our results highlight the need for complementary sky coverage between northern and southern surveys to enable robust anomaly tests in next‑generation CMB experiments.

A further conceptual consideration in anomaly studies concerns the distinction between constrained and unconstrained realizations. If the temperature features are merely statistical fluctuations consistent with the $\Lambda$CDM framework, then the $E$-mode polarization field one expects to observe is not a fully independent Gaussian realization, but rather a sky that is partially dictated by the measured temperature anisotropies, i.e., the constrained ensemble. Testing the fluke hypothesis therefore involves a direct comparison between the anomaly statistics obtained from actual polarization data and the distribution derived from constrained simulations, rather than from their unconstrained counterparts. A detection of anomalous signatures directly in the polarization data would disfavor the fluke hypothesis.

Previous investigations \cite{Shi:2022hxc,LiteBIRD:2025tnn} have concluded that, when restricted to $E$-mode only anomaly estimators, constrained and unconstrained realizations yield statistically indistinguishable outcomes, and the correlation coefficients between $T$- and $E$-mode estimators are found to satisfy $\rho\lesssim0.1$. In light of these findings, we have adopted the simpler approach of relying exclusively on unconstrained simulations to evaluate the anticipated sensitivity of AliCPT and AliCPT+SO to large-scale $E$-mode anomalies. We leave considerations of constrained realizations to future work.

\appendix                  

\section{Construction of confidence masks}
\label{app:con-msk}
To determine the AliCPT confidence masks, we have referred to the construction method of \textit{Planck} confidence masks, detailed in \cite{planck:2018yye} and Appendix A.5 of \cite{Planck:2019evm}. The \textit{Planck} common confidence mask is constructed by thresholding the standard deviation of the CMB maps between the four component-separation methods at a resolution of 80$'$, combined with the Commander, SEVEM, and SMICA confidence masks. The NILC mask is excluded because it removes a significantly smaller fraction of the sky than others.

\begin{figure}[tbp]
    \centering
    \begin{subfigure}[b]{0.54\textwidth}
        \centering
        \includegraphics[width=\textwidth]{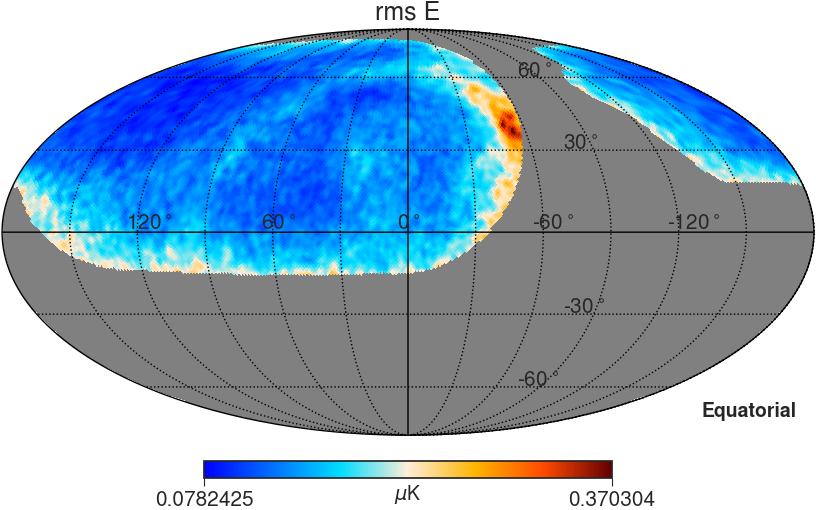}
    \end{subfigure}
    \begin{subfigure}[b]{0.45\textwidth}
        \centering
        \includegraphics[width=\textwidth]{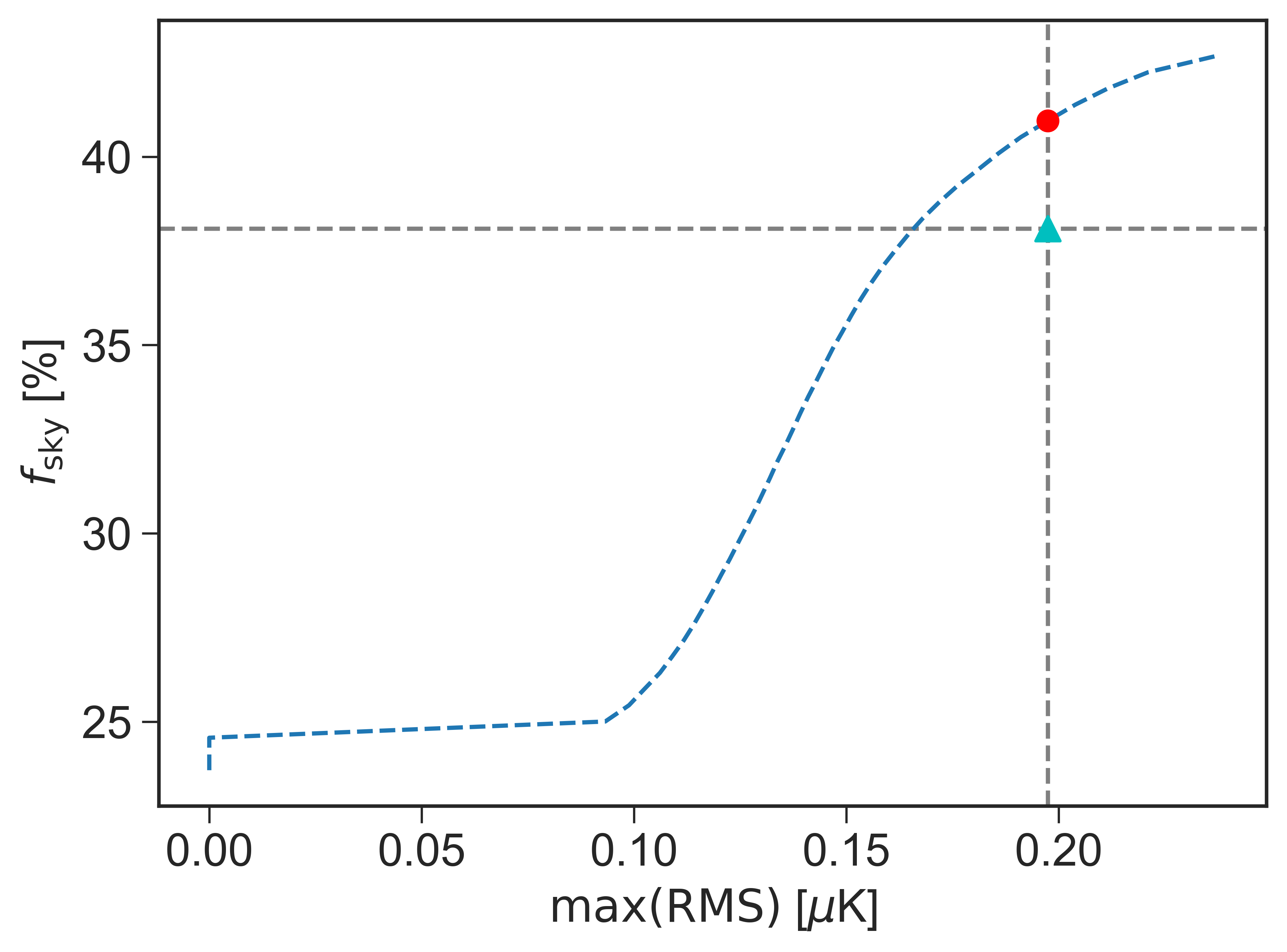}
    \end{subfigure}
    \caption{\textit{Left}: The $E$-mode RMS map (for baseline-AliCPT) representing the residual within the AliCPT wide-scan region after component separation. \textit{Right}: Maximum $E$-mode RMS residual as a function of sky fraction, for the baseline-AliCPT scenario. The red dot indicates the 0.20 $\mu$K threshold we adopt, and the cyan triangle labels the final sky fraction of the $E$-mode confidence mask.}
    \label{fig:rms_fg}
\end{figure}

In our analyses, we threshold the root mean square (RMS) map of the NILC residual between simulations to define the NILC confidence mask. The NILC pipeline is performed on $TEB$ maps derived from $IQU$ simulations with the AliCPT wide-scan mask applied. The resolution of NILC-cleaned maps is $N_{\text{side}}=64$ and $\text{FWHM}=160'$.
The root mean square residual $E$-mode map is computed by (same for temperature):
\begin{equation}
    {\rm RMS} = \sqrt{\frac{1}{N_{\rm sims}}\sum_{i=1}^{N_{\rm sims}}(\delta E)^2}\,,\quad \delta E\equiv M\cdot(\tilde{E}-E^*)\,,
    \label{eq:rms}
\end{equation}
where $M\cdot$ denotes the AliCPT wide-scan masking, $\tilde{E}$ is the NILC-cleaned $E$-mode map, and $E^*$ is the true CMB $E$-mode signal computed from the full-sky $IQU$ map. 
The residual map $\delta E$ thus contains contributions from the residual Galactic foreground contamination, the large-scale noise, and the $E/B$-mixing error due to masking effects. 
The RMS map is shown in Fig.~\ref{fig:rms_fg}. 
The Galactic mask is then defined by thresholding the RMS map at 0.20 $\mu$K (20 $\mu$K for defining the $T$-mode mask), significantly lower than the cosmological $E$-mode signal. 
The threshold is chosen as a tradeoff between reducing the errors and maximizing the sky coverage. 
Subsequently, we smooth this mask with a 5$^\circ$ FWHM Gaussian beam, reject any pixels with a value lower than 0.5 and include the remainder, in order to remove isolated small islands. The resulting binary mask with 38.1\% sky coverage is chosen as our final $E$-mode confidence mask.
When considering the NILC residual at a target sensitivity of 48 mod$\cdot$yr (instead of 4 mod$\cdot$yr), the sky coverage of the $E$-mode mask increases to 42.4\%.

We use $N_{\text{side}}=64$ maps for map-based anomaly analyses and \revise{$N_{\text{side}}=32$ maps for harmonic-based analyses. The $N_{\text{side}}=32$ masks are derived similarly to the low-resolution data, by smoothing and downgrading the $N_{\text{side}}=64$ confidence mask to 320$'$ FWHM and $N_{\text{side}}=32$, and thresholding the smoothed mask at a value of 0.9. The final sky fraction of the $N_{\text{side}}=32$ $E$-mode mask is 34.9\%} (see Table~\ref{tab:fsky}).

\section{The form of dipole modulation in harmonic space}
\label{app:dm-harm}

The dipole modulation defined by Eq.~\eqref{eq:dm} can be recast in harmonic space as \cite{Zibin:2015ccn,Contreras:2017zjv}:
\begin{equation}
    a_{\ell m}=\tilde a_{\ell m}+\sum_{M=0,\pm1}\Delta X_M\sum_{\ell 'm'}\tilde a_{\ell'm'}\xi^M_{\ell m\ell'm'}\,,
    \label{eq1.2}
\end{equation}
where $a_{\ell m}$ and $\tilde a_{\ell m}$ denote spherical harmonic coefficients of the modulated and isotropic CMB, respectively. $\Delta X_M$ is defined as the spherical harmonic decomposition of the dipole, $\boldsymbol d=(A_d, \theta_d, \phi_d)$ in spherical coordinates, with the specific forms: 
\begin{equation}
    \begin{aligned}
    \Delta X_0&=A_d\cos\theta_d\,,\\
    \Delta X_{+1}&=-\frac{A_d}{\sqrt{2}}\sin\theta_d e^{-i\phi_d}\,,\\
    \Delta X_{-1}&=\frac{A_d}{\sqrt{2}}\sin\theta_d e^{i\phi_d}=-\Delta X_{+1}^*\,.
    \end{aligned}
\end{equation}
The coupling coefficients $\xi^M_{\ell m\ell'm'}$ are defined as:
\begin{equation}
    \xi^M_{\ell m\ell'm'} \equiv \sqrt{{\frac{4\pi}{3}}}\int Y_{\ell'm'}(\hat{\boldsymbol n})Y_{1M}(\hat{\boldsymbol n})Y^*_{\ell m}(\hat{\boldsymbol n})d\Omega_{\hat{\boldsymbol n}}\,,
    \label{eq1.5}
\end{equation}
where $Y_{\ell m}(\hat{\boldsymbol n})$ are spherical harmonic functions.
More explicitly,
\begin{equation}
\begin{aligned}
        &\xi^0_{\ell m\ell'm'}=\delta_{m'm}(\delta_{\ell'\ell-1}A_{\ell -1m}+\delta_{\ell'\ell+1}A_{\ell m})\,,
    \\
    &\xi^{\pm1}_{\ell m\ell'm'}=\delta_{m'm\mp1}(\delta_{\ell'\ell-1}B_{\ell -1\pm m-1}-\delta_{\ell'\ell+1}B_{\ell \mp m})\,,
\end{aligned}
\end{equation}
where
\begin{equation}
\begin{aligned}
        &A_{\ell m}=\sqrt{\frac{(\ell+1)^2-m^2}{(2\ell+1)(2\ell+3)}}\,,
\\
    &B_{\ell m}=\sqrt{\frac{(\ell+m+1)(\ell+m+2)}{2(2\ell+1)(2\ell+3)}}\,.
\end{aligned}
\end{equation}

For polarization, it is convenient to use relations \cite{Ghosh:2015qta,Ghosh:2018apx}:
\begin{equation}\label{eq:mod-alm-eb}
\begin{aligned}
        &a_{\ell m}^E=\tilde a^E_{\ell m}+A_d\alpha_{-}\tilde a^E_{\ell -1m}+iA_d\alpha_{0}\tilde a^B_{\ell m}+A_d\alpha_{+}\tilde a^E_{\ell +1m}\,,
\\
    &a_{\ell m}^B=\tilde a^B_{\ell m}+A_d\alpha_{-}\tilde a^B_{\ell -1m}-iA_d\alpha_{0}\tilde a^E_{\ell m}+A_d\alpha_{+}\tilde a^B_{\ell +1m}\,,
\end{aligned}
\end{equation}
where
\begin{equation}
\begin{aligned}
        &\alpha_{-}=\frac{1}{\ell}\sqrt{\frac{(\ell-2)(\ell+2)(\ell-m)(\ell+m)}{(2\ell-1)(2\ell+1)}}\,,
\\
    &\alpha_0=\frac{2m}{\ell(\ell+1)}\,,
    \\
    &\alpha_{+}=\frac{1}{\ell+1}\sqrt{\frac{(\ell-1)(\ell+3)(\ell-m+1)(\ell+m+1)}{(2\ell+1)(2\ell+3)}}\,.
\end{aligned}
\end{equation}
Here the $E$- and $B$-mode spherical harmonic coefficients are rotated such that the $z$-axis aligns with the dipole ${\boldsymbol d}$, then modulated using Eq.~\eqref{eq:mod-alm-eb}, and finally derotated backward.

\section{\revise{Justification of downgrading}}
\label{app:just-repix}

\begin{figure}[tbp]
    \centering
    \begin{subfigure}[b]{0.49\textwidth}
        \centering
        \includegraphics[width=\textwidth]{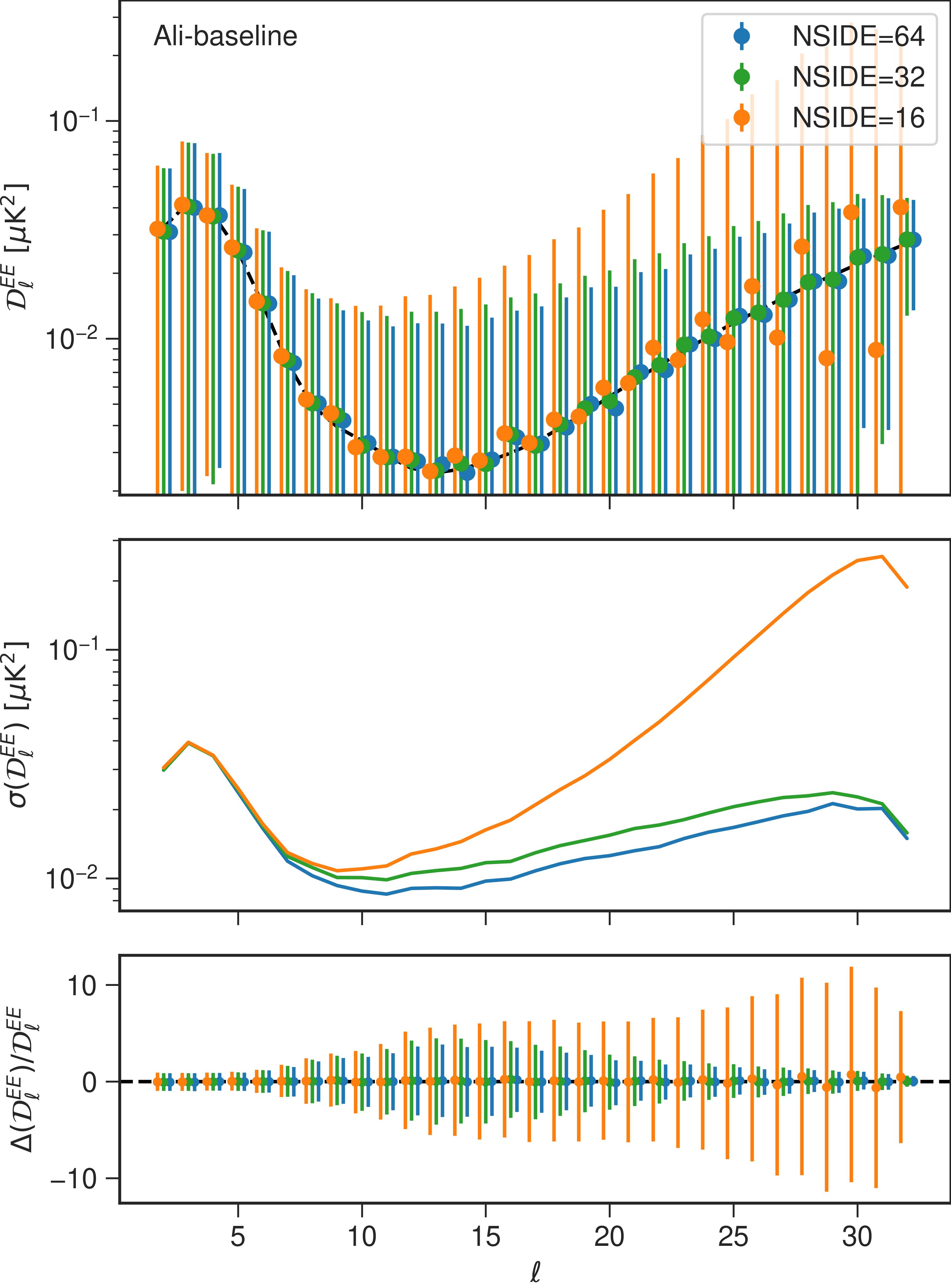}
    \end{subfigure}
    \begin{subfigure}[b]{0.49\textwidth}
        \centering
        \includegraphics[width=\textwidth]{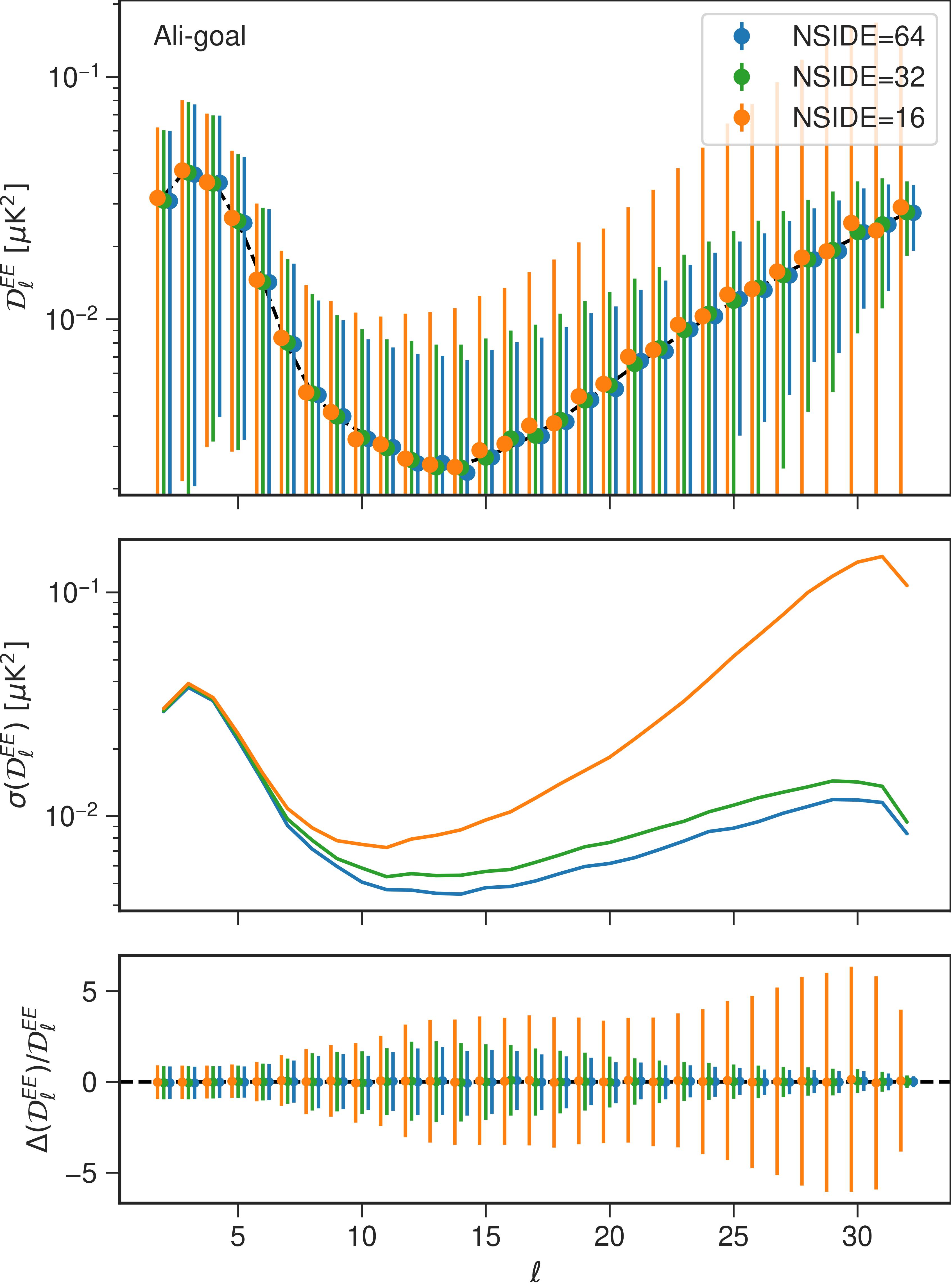}
    \end{subfigure}
    \\
    \begin{subfigure}[b]{0.4\textwidth}
        \centering
        \includegraphics[width=\textwidth]{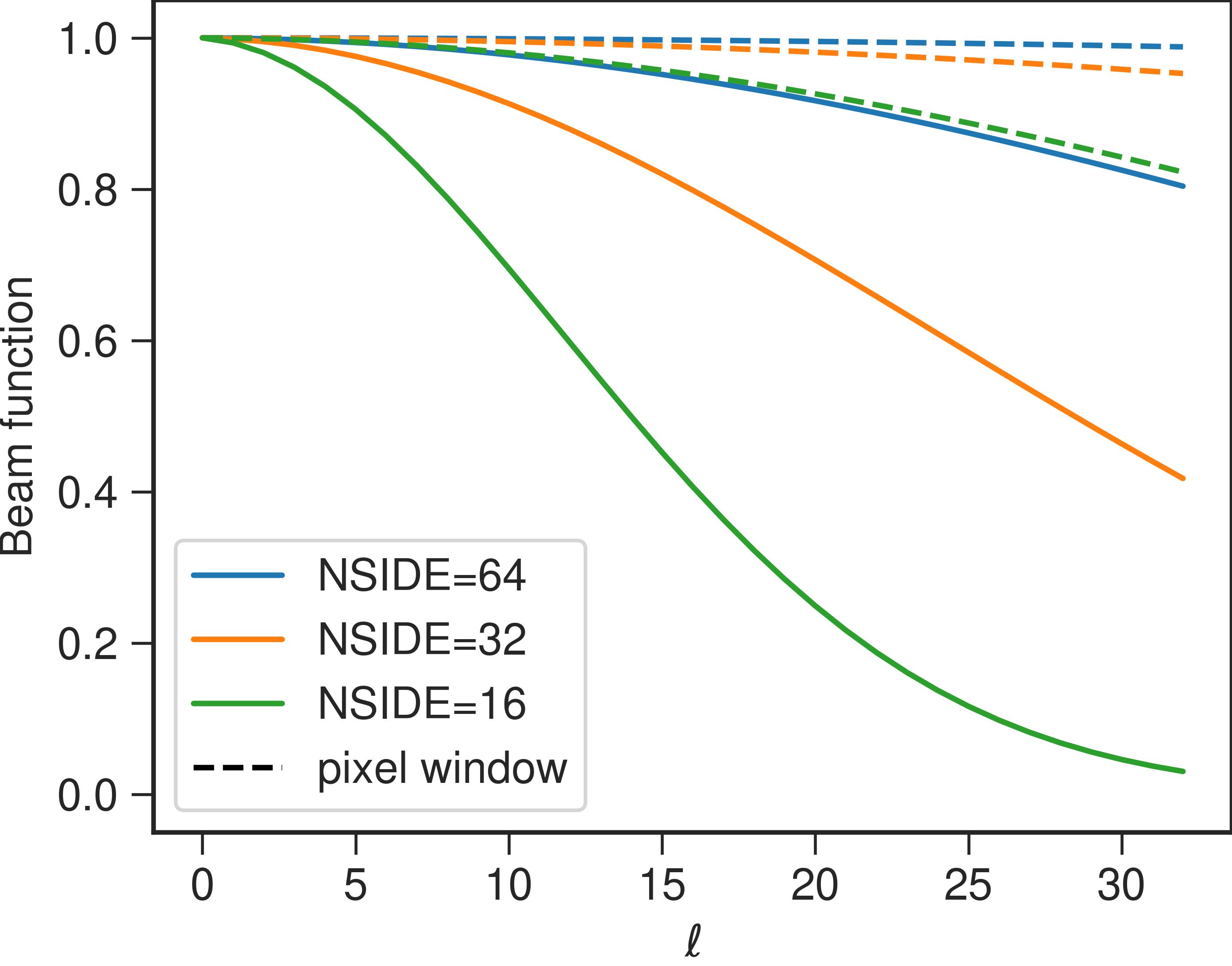}
    \end{subfigure}
    \caption{\textit{Upper}: \texttt{xQML} $EE$ power spectra computed from NILC maps with different \texttt{HEALPix} resolutions, for baseline-AliCPT (left) and goal-AliCPT (right) cases (as plotted in Fig.~\ref{fig:dl-xqml}). \textit{Lower}: Beam transfer functions (solid curves) and pixel window functions (dashed curves) for the same \texttt{HEALPix} resolutions.}
    \label{fig:compar-nside}
\end{figure}

\revise{Since computing \texttt{xQML} power spectra from $N_{\text{side}}=64$ maps is prohibitively expensive, we downgrade the NILC maps to $N_{\text{side}}=32$ and apply a 320$'$ Gaussian beam smoothing. Here, we demonstrate that the downgrading process introduces almost no bias or additional uncertainty in the estimated $EE$ spectra. Fig.~\ref{fig:compar-nside} presents the \texttt{xQML} $EE$ power spectra computed from input maps at $N_{\text{side}}=64$, 32, and 16, for both the AliCPT baseline and goal noise cases. The spectra at $N_{\text{side}}=32$ closely match those at $N_{\text{side}}=64$. However, the $N_{\text{side}}=16$ results show substantially larger errors at $\ell\gtrsim20$, where the corresponding beam transfer function decays to zero, making this resolution unsuitable for our analysis. These findings justify the repixelization to $N_{\rm side}=32$ for harmonic-based $E$-mode anomaly analysis of AliCPT.}

\acknowledgments
This work is supported by the National Key R\&D Program of China Grant No. 2021YFC2203102, NSFC No. 12325301 and 12273035.

\bibliography{Bib}
\bibliographystyle{JHEP}

\end{document}